\documentclass[sn-mathphys-num]{sn-jnl}

\usepackage{graphicx}%
\usepackage{multirow}%
\usepackage{amsmath,amssymb,amsfonts}%
\usepackage{amsthm}%
\usepackage{mathrsfs}%
\usepackage[title]{appendix}%
\usepackage{textcomp}%
\usepackage{manyfoot}%
\usepackage{booktabs}%
\usepackage{algorithm}%
\usepackage{algorithmicx}
\usepackage{algpseudocode}%
\usepackage{listings}%
\usepackage[utf8x]{inputenc}
\usepackage{graphicx}
\usepackage{threeparttable}
\graphicspath{{images/}}
\usepackage{parskip}
\usepackage{fancyhdr}
\usepackage{vmargin}
\usepackage{cite}      
\usepackage{psfrag}    
\usepackage{subfigure}
\usepackage{bm} 
\usepackage{mathtools}
\usepackage{stfloats} 
\usepackage{multicol}
\usepackage{comment}
\usepackage{float}

\usepackage{enumitem}

\newcommand{\beq}{\begin{equation}}
\newcommand{\eeq}{\end{equation}}
\newcommand{\beqa}{\begin{eqnarray}}
\newcommand{\eeqa}{\end{eqnarray}}

\begin{document}

\title[Sequential Filtering Techniques for Simultaneous Tracking and Parameter Estimation]{Sequential Filtering Techniques for Simultaneous Tracking and Parameter Estimation}


\author*[1,2]{\fnm{Yannick} \sur{Sztamfater Garcia}}\email{ysztamfa@pa.uc3m.es}

\author[2]{\fnm{Joaquin} \sur{Miguez}}\email{jmiguez@ing.uc3m.es}

\author[1,3]{\fnm{Manuel} \sur{Sanjurjo Rivo}}\email{msanjurj@ing.uc3m.es}

\affil*[1]{\orgdiv{Department of Aerospace Engineering}, \orgname{Universidad Carlos III of Madrid}, \orgaddress{\street{Avenida de la Universidad, 30}, \city{Leganés}, \postcode{28911}, \country{Spain}}}

\affil[2]{\orgdiv{Department of Signal Theory and Telecommunications}, \orgname{Universidad Carlos III of Madrid}, \orgaddress{\street{Avenida de la Universidad, 30}, \city{Leganés}, \postcode{28911}, \country{Spain}}}

\affil[3]{\orgname{Northstar Earth and Space}, \city{Luxembourg}, \country{Luxembourg}}


\abstract{The number of resident space objects is rising at an alarming rate. Mega-constellations and breakup events are proliferating in most orbital regimes, and safe navigation is becoming increasingly problematic. It is important to be able to track RSOs accurately and at an affordable computational cost. Orbital dynamics are highly nonlinear, and current operational methods assume Gaussian representations of the objects' states and employ linearizations which cease to hold true in observation-free propagation. 
Monte Carlo-based filters can provide a means to approximate the a posteriori probability distribution of the states more accurately by providing support in the portion of the state space which overlaps the most with the processed observations. Moreover, dynamical models are not able to capture the full extent of realistic forces experienced in the near-Earth space environment, and hence fully deterministic propagation methods may fail to achieve the desired accuracy. By modeling orbital dynamics as a stochastic system and solving it using stochastic numerical integrators, we are able to simultaneously estimate the scale of the process noise incurred by the assumed uncertainty in the system, and robustly track the state of the spacecraft. In order to find an adequate balance between accuracy and computational cost, we propose three algorithms which are capable of tracking a space object and estimating the magnitude of the system's uncertainty. The proposed filters are successfully applied to a LEO scenario, demonstrating the ability to accurately track a spacecraft state and estimate the scale of the uncertainty online, in various simulation setups.}

\keywords{Tracking, sequential filters, particle filters, SMC, parameter estimation, uncertainty propagation}



\maketitle
\newpage
\section{Introduction}
\subsection{Orbit determination}
The last decade has seen a surge in resident space objects (RSOs) as a result of the deployment of large mega-constellations and the increase in launch capabilities, both from the public and the private sectors. As of November 2024, there are over 34,000 catalogued objects currently orbiting the Earth, out of which just over 12,000 are operational spacecraft, the remainder including long de-serviced satellites, rocket parts and space debris \citep{ESAdata}. There exist entities responsible for the tracking and cataloging of RSOs to provide space surveillance and tracking (SST) services such as reentry or conjunction assessment. The data processing core of their operations is based on orbit determination (OD) methods which are able to provide orbital state and uncertainty estimates given observational data. Although the target tracking problem has a long history in many different applications, OD is challenging due to its unique characteristics. Indeed, methods aimed at estimating the state of an RSO from measurements of radar, electro-optical sensors, laser or other sensors (generally not on-board sensors) have to deal with sparse data and non-linear dynamical and measurement models. 

OD methods can be classified (see Figure \ref{fig:filters_hier}) in three broad categories:
\begin{itemize}
    \item batch least-squares filters\citep{Gelb}, where all observations are processed together, yielding a solution through iterative fitting,
    \item sequential filters \citep{Maybeck}, where the orbital model parameters are updated with each new observation; 
    \item recursive least squares, or batch sequential filters (BSF) \citep{Fraser}\citep{Patil}, which combine batch least squares and sequential filter properties.
\end{itemize} 
In this work, the focus is on sequential filtering because of their potential to yield efficient online OD. More specifically, within sequential filters, we identify three popular families: Gaussian filters (starting with \citep{Kalman}), particle filters \citep{Gordon} and hybrid particle filters (\citep{Alspach}\citep{Doucet2000}), depending on the way the prediction (in which the prior estimate is propagated to the next observation) and update steps (in which the prior estimate is corrected with new observational data) are performed. 

\subsection{Filtering algorithms}
Numerous filtering techniques have been proposed over the last 60 years to deal with the problem of state estimation in various fields.
The Kalman filter (KF) \citep{Kalman} is designed to provide an optimal solution to linear systems with additive Gaussian noise. Non-linear systems, however, must be tackled through approximate methods such as the extended Kalman filter (EKF) \citep{Anderson}, the unscented Kalman filter (UKF) \citep{Julier} and the cubature Kalman filter (CKF) \citep{ckf}. The EKF, which introduces linearizations for the transition or observation functions is compared with a least squares algorithm (LSQ) in \citet{Segan}, demonstrating that although LSQ is more robust, the EKF is more efficient in real-time. The UKF and the CKF propagate a set of sigma (or cubature) points, improving probability distribution representation after non-linear transformations and bypassing the need for linearizations. CKF methods include a third-degree flavour (3CKF), comparable to the UKF, and the fifth-degree flavour (5CKF), introduced by \citet{5ckf} for a real-time low-Earth orbit scenario and shown to provide higher accuracy than the 3CKF. In all Kalman filters, the Kalman gain matrix maps residuals to corrections of the previous estimate, balancing the weight of observational data and model dynamics. 

%

Unlike Kalman-style methods, particle filters (PFs) can account for probability distributions which are possibly non-Gaussian. PFs are recursive Monte Carlo (MC) methods that can be used to numerically approximate the sequence of posterior probability distributions of the state given a set of observations \citep{Doucet2001}\citep{Djuric2003}\citep{Cappe}. The basic theory of particle filtering deals with dynamic variables only; hence all other parameters are assumed to be known. The simplest form of PF is the bootstrap particle filter (BPF) (see \citep{Gordon}\citep{Kitagawa}), where Monte Carlo samples at each time are drawn from the state distribution given by the dynamical model, i.e., the probability distribution (termed "the proposal" in PFs) associated to the propagated samples. However, customized proposals can be constructed to improve convergence \citep{Maskell}. Particles are then allocated importance weights according to their likelihood. Resampling of the particles using these weights is a fundamental step in particle filtering \citep{Djuric15}. However, weight degeneracy (a phenomenon which occurs when one single particle accumulates most of the weight \citep{Snyder08} can easily lead to sample impoverishment after the resampling step. \citet{Pardal1} analyzes sample impoverishment in PFs, and proposes strategies like regularization and resampling thresholds to mitigate this issue. Resampling is not needed at every time step; one can choose to resample once a threshold is reached. This is often done by evaluating the effective sample size, which is an estimate of the variance of the (non-normalized) weights \citep{Cappe}. The downsides associated with weight degeneracy can often be challenging to mitigate without incurring a large computational effort. \citet{McCabe} reviews various PF implementations for space object tracking, emphasizing adaptability to non-Gaussian uncertainties and multi-modal distributions. They demonstrate significant performance advantages over the UKF in different orbital scenarios. 
\citet{Mashiku} employs a PF to model uncertainties that deviate from Gaussianity in OD, highlighting significant improvements in accuracy (with errors around $25\%$ lower than an EKF) in cases with large initial uncertainties, while \citet{Escribano} uses a PF where each particle represents a possible spacecraft state under various maneuver hypotheses, allowing the algorithm to probabilistically infer the most likely maneuver scenario. 

The ensemble Kalman Filter (EnKF)\citep{Evensen} combines Monte Carlo sampling with Kalman updates, to construct a random ensemble rather than deterministic sigma points, making it popular in high-dimensional problems. It aims at achieving substantial coverage of the state space, whilst propagating equally-weighted particles and, hence, not requiring sample-weighing and resampling steps. It is therefore often more robust than PFs in cases where the latter suffers from weight degeneracy.
\citet{Gamper} apply an EnKF to spacecraft tracking in LEO with simulated observations and a covariance inflation parameter, though achieving errors in position considerably larger than a UKF. 
\citet{DeMars_GMM} applies Gaussian mixture models (GMM) to initial orbit determination (IOD), enabling probabilistic representations of uncertainties, while \citet{Yun1} introduces a kernel-based Gaussian mixture filter that combines ensemble sampling with non-parametric density estimation. These approaches significantly improve robustness and accuracy with limited data. Other hybridizations include works by \citet{Chakravorty2}, which combine the UKF with a PF to manage non-Gaussian uncertainties, achieving a balance between computational efficiency and estimation accuracy which outperforms the standalone UKF and PF by up to $40\%$ in tracking maneuvers in simulated scenarios.

\begin{figure}[H]
    \centering
    \includegraphics[scale=0.42]{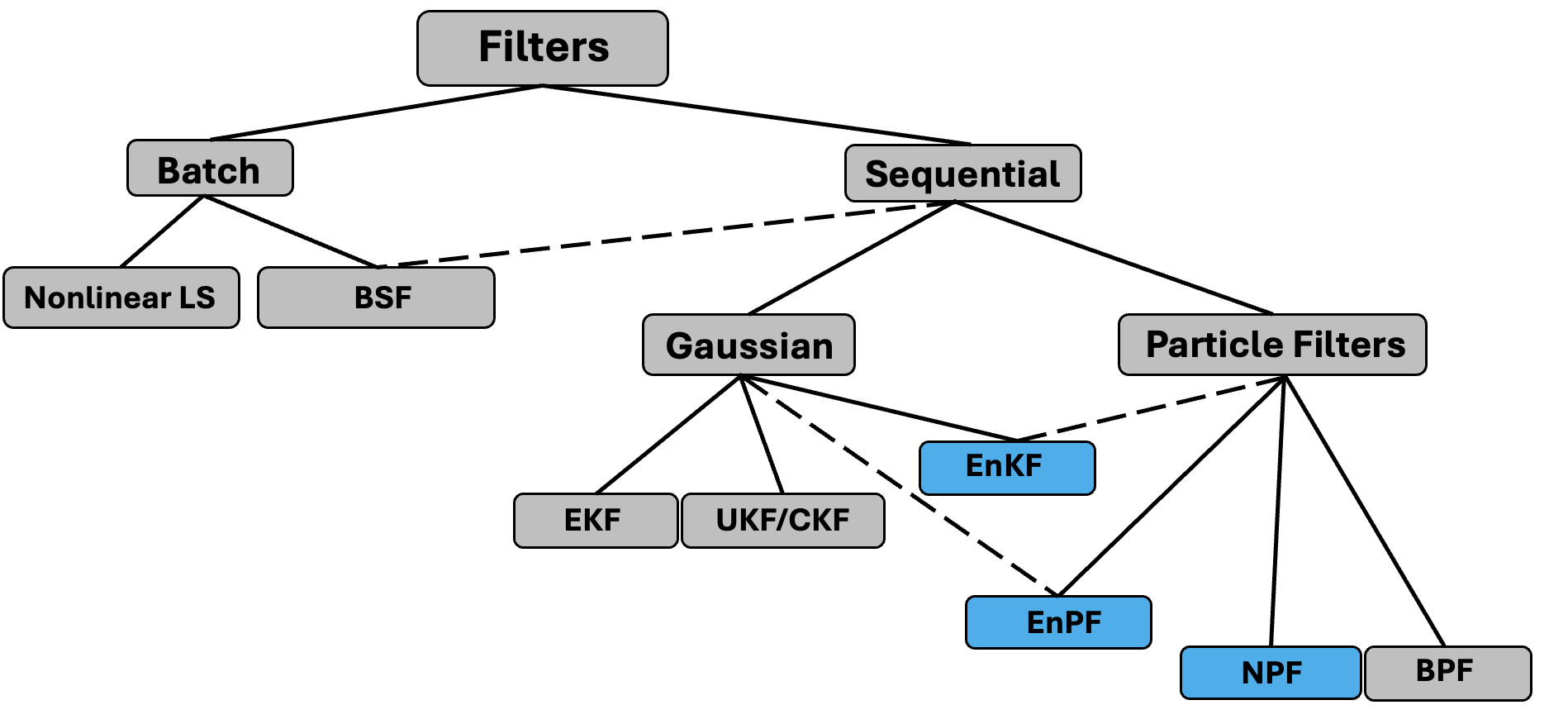}
    \caption{Hierarchy of the main types of filters used in OD and RSO tracking. The main groups are batch and sequential methods, each spanning their sub-types (gray boxes). The filtering methods proposed in this work are shown in light blue boxes. }
    \label{fig:filters_hier}
\end{figure}

Figure \ref{fig:filters_hier} shows the hierarchy of the most relevant OD methods discussed above, as well as the novel algorithms introduced in this paper (in blue).

\subsection{Characterization of process noise}
For all OD methods, the estimation performance depends on the correct characterization of the process and measurement noise. In fact, the development of optimal estimators in the Bayesian sense (i.e. those that solve the problem of the exact and complete characterization of the a posteriori probability distribution) is based on the knowledge of suitable models for the dynamical and measurement noise. In the case of KFs for linear time invariant systems, the identification of noise covariances is a problem that has been studied for 50 years \citep{zhang2020identification}, and is not yet fully solved.
The problem of dealing with inexact knowledge of the noise present in the model can be formulated as follows: 
Given a vector time series and a library of models of system dynamics, find a suitable process and measurement noise model and the best system dynamics for the time series. The traditional classification of methods for dealing with this problem is summarized in Table \ref{tab:methods}, where MPE stands for measurement prediction error.

\begin{table}[h]
\caption{Methods for noise characterization.} 
\normalsize
        \begin{tabular}{|p{0.3\linewidth} | p{0.3\linewidth} | p{0.3\linewidth}|}
        \hline
         &  \bf{State and noise simultaneous estimation} & \bf{Noise-only estimation}\\
        \hline 
        \bf{Probability-based}      &  Bayesian inference  & Maximum likelihood estimation \\
        \hline
        \bf{Statistical Analysis of MPE} & Covariance Matching Method (CMM) & Correlation methods \\
        \hline
        \end{tabular}
           \label{tab:methods}
\end{table}

Methods that deal with noise characterization alone are based on the solution of a sub-optimal state estimator, whilst algorithms that deal with noise and state estimation simultaneously do so by increasing the state dimension with parameters that allow the noise to be characterized in some way. Note that noise can be characterized, but not estimated. It is possible, however, to estimate the variance of noise or the diffusion coefficient of a stochastic differential equation.
It has been shown that estimators in this type of problem are necessarily nonlinear \citep{dunik2017noise}, even in the case of linear dynamic and measurement models, because of the nonlinear relationship between the elements of the extended state. It is therefore necessary to use nonlinear filters (such as PFs or hybrid schemes), in this type of problem. 
Specifically, in this work, we focus on the Bayesian inference approach in which noise estimation is performed by approximating the posterior probability distribution of any unknown parameters within the framework of hybrid filtering. 
Moreover, the inclusion and study of process noise in orbital dynamics has received relatively little attention in the astrodynamics community \citep{poore2016covariance}. This process noise is commonly simply accounted for by scaling or inflating its covariance matrix throughout the propagation, but the magnitude of the scaling is often rather arbitrary. 
To achieve realistic covariance estimates, refined uncertainty quantification techniques are essential. Methods such as those proposed in \citep{Stacey1}, which present innovation-based approaches for updating noise covariance during filtering over a sliding time window, or in \citet{Stacey2}, which incorporate empirical accelerations modeled as a first-order Gauss-Markov process in KF variations, involve estimating and incorporating uncertainties from either observations or dynamical models into the OD process, enhancing the accuracy of the filter. Other works achieve this by employing consider parameter analysis; e.g., \citet{Cano1} and \citet{Cano2} extend the process noise covariance matrix to more appropriately represent uncertainty in the time update phase. 

\subsection{Contributions}
In this paper, we propose a set of hybrid Monte Carlo filters for the simultaneous tracking of spacecraft states and characterization of process noise, enabled by the utilization of Itô stochastic differential equations (SDEs) to model orbital dynamics under uncertainty, and hence allowing us to estimate the noise parameters in the SDE.
Orbital dynamics are commonly modeled deterministically by way of ordinary differential equations (ODEs). Uncertainty due to both aleatory and epistemic sources can be taken into account by employing a stochastic parametrization of the unknown perturbations and accelerations affecting the orbit. This can be done by using an Itô SDE instead of an ODE \citep{luo}. SDEs involve a drift term, which is deterministic and given by the known accelerations, as well as a stochastic term. The latter is parametrized by a diffusion coefficient that accounts for the scale of the unknown perturbations. This parameter can be estimated by employing the three types of hybrid Monte Carlo filters herein proposed, most of which have not, to the best of our knowledge, been used in astrodynamics before. 
The stochastic parametrization of the unknown perturbations enables the proposed filters to accurately target the state of an operational spacecraft using simplified models.
The results are comparable with industry standards and are achieved with a tractable computational cost.

Section \ref{sec:ssd} provides background on the Bayesian filtering process and introduces the state space model of interest. In Section \ref{sec:methodology} we present the standard PF as a reference, and a set of hybrid Monte Carlo filters for joint state tracking and noise calibration. The algorithms include an EnKF-like method, a novel, greedy PF that combines Kalman updates with importance weights, and a nested hybrid filter (NHF), in the vein of \citep{PerezVieites2021}. After defining the test cases in Section \ref{sec:test_case}, Sections \ref{sec:results1} and \ref{sec:results2} are devoted to the presentation and discussion of the numerical results for the various filters. Some concluding remarks are presented in Section \ref{sec:CONC}.

\section{State space definition}\label{sec:ssd}
\subsection{Dynamical and observation models}\label{sec:dynmod}
The aim is to track an orbiting spacecraft by making use of sequential filters. Hereafter, we use regular-face letters to represent scalars (e.g., $x \in \mathbb{R}$), bold-face lower-case letters for column vectors (e.g., ${\bf x} = [r_x, r_y, r_z, v_x, v_y, v_z]^\top$) and bold-face upper-case letters for matrices (e.g., ${\bf X} = [{\bf x}(0), {\bf x}(1)]$, where ${\bf x}(0)$ and ${\bf x}(1)$ are column vectors of the same dimension).
Let ${\bm { x}}(t)  = \begin{bmatrix} \bm{{ r}}(t) \\ \bm{{ v}}(t) \end{bmatrix}$ denote the 6-dimensional state of the spacecraft at time $t$, where ${\bm r}(t)$ denotes position and ${\bm v}(t)$ is velocity.
Assume a Gaussian prior distribution at $t_0$, $p(\bm{x}(t_0)) \equiv \mathcal{N}({\bm {\hat x}}(t_0), {\bm \Sigma}_0)$, given by mean ${\bm {\hat x}}(t_0)  = \begin{bmatrix} \bm{{\hat r}}(t_0) \\ \bm{{\hat v}}(t_0) \end{bmatrix}$, and covariance matrix $\mathbf{\Sigma}_{0}$, implying that a previous OD procedure has been performed.
Orbital dynamics are commonly modeled as a deterministic system of equations of motion. 
The dynamical model evaluation is then given by 
\begin{equation}
    \frac{\text{d}{\bm x}}{\text{d}t}=f(\bm x,t) :=
    \begin{bmatrix}
        \bm{v} \\
        -\frac{\bar\mu \bm{r}}{r^3}  + \bm{a}_{\text{pert}}
    \end{bmatrix}
    \label{eq:rdot}
\end{equation}
where $-\frac{\bar\mu \bm{r}}{r^3}$ represents the two-body gravitational acceleration, and $\bm{a}_{\text{pert}}$ represents perturbative accelerations to two-body motion. However, this deterministic approach does not account for inaccuracies in the dynamical model or random errors that accumulate during propagation. To address this problem, we adopt a stochastic model that incorporates a noise term to represent propagation uncertainty \citep{luo}. Eq. \eqref{eq:rdot} can be extended into a stochastic model where $f(\bm x,t)$ becomes the drift of the stochastic state $\bm x(t)$ and a diffusion (noise) term driven by a $d$-dimensional Wiener process $\bm W(t)$ is introduced. Specifically, we convert Eq. \eqref{eq:rdot} into the Itô SDE
\begin{equation}\label{eq:sde}
\text{d}\bm{x} = f(\bm{x},t)\text{d}t + {\bm\sigma_W}(\bm{x})\text{d}{W},
\end{equation}
where ${\bm\sigma_W}(\bm{x})$ is a 6-dimensional diffusion coefficient matrix, and we assume the process noise is only present in the velocity components. This matrix is given by
\begin{equation*}
\bm{\sigma}_W(\bm x) = \bm\sigma
\begin{bmatrix}
\mathbf{0}_{3 \times 3} & \mathbf{0}_{3 \times 3} \\
\mathbf{0}_{3 \times 3} & \mathbb{I}_{3\times 3}
\end{bmatrix}
\end{equation*}
where $\bm \sigma = [0,\ 0,\ 0,\ \sigma_{W}({v_x}),\ \sigma_{W}(v_y),\ \sigma_{W}(v_z)]^\top$. 
This matrix captures the uncertainty in the dynamical model accelerations due to unmodeled dynamics or error. The stochastic parametrization of the unknown perturbations provides a more realistic characterization of spacecraft motion under uncertainty.

Tracking and prediction of the state $\bm x(t)$ is usually carried out by processing observations collected at certain time instants. The types of observations depend on the orbital region and availability of the sensors. For the time being, we assume a general (abstract) model in which an observation at time $t$ is represented as
\begin{equation}\label{eq:obs}
    {\bm z}(t) = \mathcal{M}({\bm x},t) + \bm s(t),
\end{equation}
where $\mathcal{M}(\cdot,t)$ is the transfer function of the sensor(s) available at time $t$  and $\bm s(t)$ is observational noise, which we model as Gaussian, with zero mean and covariance matrix $\Omega(t)$, i.e., $\bm s(t) \sim \mathcal{N}(0,\Omega(t))$. 

In low-Earth orbit (LEO), observations may come from ground-based radar sensors, and are typically comprised of range, range rate, azimuth, and elevation exclusively. High accuracy in orbit estimates requires precise measurement models, which account for geometric and dynamic factors and corrections to incorporate environmental effects. 


\subsection{Numerical integration of the dynamical model}\label{sec:numscheme}
Assume a time grid ${\bm T} = \{t_l\}_{l=0}^L $, where  $t_0$ is the initial time and $t_1, \dots, t_L$ are time instants at which the value of the state $\bm x(t)$ must be approximated.

The Itô SDE in Eq. \eqref{eq:sde} must be discretized using a stochastic numerical scheme with a step-size $h_l = t_l-t_{l-1} > 0$.
Due to its simplicity, the Euler-Maruyama discretization scheme is described as
\begin{equation}
\label{eq:eulersde}
    {\bm {\tilde x}}_l = \tilde{\bm x}_{l-1} + f(\tilde{\bm x}_{l-1},t)h_l+{\bm\sigma}_{W}(\tilde {\bm x}_{l-1})\bm{W}_{l}
\end{equation}
where $l \in \mathbb{N}$ is discrete time, $\tilde {\bm x}(t)$ is the approximate state at time $t_l$, $\tilde {\bm x}_l \approx \bm x(t_l)$
and $\bm W_l=\bm W(t_l)-\bm W(t_{l-1})$ is a Gaussian d-dimensional r.v. with zero mean and diagonal covariance matrix $h_l\mathbb{I}$. The stochastic diffusion coefficient is assumed independent of ${\bm x}$, i.e., ${\bm\sigma}_{W}({\bm x}_l) = {\bm\sigma}_{W}$
In principle, any stochastic discretization scheme can be used. However, for sufficient precision, Euler schemes in orbital dynamics demand very low step-sizes, which increase the computational cost.
In this work, the stochastic Runge-Kutta (SRK) integration scheme of \citet{Rumelin1982} is used, which is a well-known stochastic adaptation of the classical Runge-Kutta of order 4 (RK4), though with strong order 1 convergence \citep{Kloeden}. 


Numerical schemes enable us to (approximately) sample the state at any given time instants. For example, Eq. \eqref{eq:eulersde} determines the conditional probability density function (pdf) of $\tilde {\bm x}_l \approx \bm x(t_l)$ given $\tilde {\bm x}_{l-1} \approx \bm x(t_{l-1})$. Specifically, we denote this conditional pdf as $p(\tilde {\bm x}_l|\tilde {\bm x}_{l-1})$ and it is easy to see that 
\begin{equation*} 
    p(\tilde {\bm x}_l|\tilde {\bm x}_{l-1}) = \mathcal{N}(\tilde {\bm x}_l;\tilde {\bm x}_{l-1}+h_lf(\tilde {\bm x}_{l-1},t_{l-1}),\sigma_W\sigma_W^\top).
\end{equation*}
If other schemes are used, the expression for $p(\tilde {\bm x}_l|\tilde {\bm x}_{l-1})$ may become more involved but sampling $\tilde {\bm x}_l\sim p(\tilde {\bm x}_l|\tilde {\bm x}_{l-1})$ is always possible because it amounts to running one step of the numerical scheme at hand. 
Moreover, for any $m>0$ the conditional pdf $p(\tilde {\bm x}_{l+m}|\tilde {\bm x}_l)$ can be expressed as 
\begin{equation*}
    p(\tilde {\bm x}_{l+m}|\tilde {\bm x}_{l}) = \int \ldots\int \prod_{i=1}^m p(\tilde {\bm x}_{l+i}|\tilde {\bm x}_{l+i-1})\text{d}\bm x_{l+i}\ldots \text{d}\bm x_{l+m-1},
\end{equation*}
and sampling $\tilde {\bm x}_{l+m}\sim p(\tilde {\bm x}_{l+m}|\tilde {\bm x}_l)$ amounts to running $m$ steps of the numerical scheme starting at $\tilde {\bm x}_l$.

\subsection{Bayesian filtering}
Assume that observations are collected at times $\bm T' = \{t'_k\}_{k=1}^M$ and denoted $\bm z_k=\bm z(t'_k)$ for conciseness. Also assume that the numerical scheme of Section \ref{sec:numscheme} is aligned with the observation times in the sense that there are integer indices $l_1,\ldots,l_M$ such that $t'_k = t_{l_k}$. Furthermore, denote $\bm x_k = \tilde {\bm x}_{l_k}$, hence, $\bm x_k \approx \bm x(t'_k)$. 
Given the observation model of Eq. \eqref{eq:obs}, the conditional pdf of the observation $\bm z_k$ given the state $\bm x_k$ is 
\begin{equation*}
    p(\bm z_k|\bm x_k) = \mathcal{N}(\bm z_k;\mathcal{M}_k(\bm x_k,t'_k),\Omega_k),
\end{equation*} 
i.e., it is Gaussian and it can be evaluated for any value of $\bm x_k$. 
We also note that the conditional pdf of $\bm x_k$ given $\bm x_{k-1}$ is
\begin{equation*}
    p({\bm x}_{k}|{\bm x}_{k-1}) = p(\tilde {\bm x}_{l_k}|\tilde {\bm x}_{l_{k-1}}) = \int \ldots\int \prod_{i=l_{k-1}}^{l_k} p(\tilde {\bm x}_{i}|\tilde {\bm x}_{i-1})\text{d}\tilde {\bm x}_{l_{k-1}+1}\ldots \text{d}\tilde{\bm x}_{l_k-1},
\end{equation*}
hence, we can sample $\bm x_k$ from $\bm x_{k-1}$ simply taking $l_k-l_{k-1}$ steps of the numerical scheme. 

From a Bayesian perspective, the statistical characterization of the state $\bm x_k$ at time $t'_k$ given the data available up to that time is contained in the a posteriori pdf $p(\bm x_k|\bm z_{1:k})$. A Bayesian filter is a recursive algorithm that computes (or, at least approximates) the sequence of pdfs $p(\bm x_k|\bm z_{1:k}),\ k=1,2,\ldots$ as the observations are sequentially collected. 
To be specific, given the filtering pdf at time $t'_{k-1}$, $p(\bm x_{k-1}|\bm z_{1:k-1})$, one can use a Chapman-Kolmogorov equation to obtain the one-step-ahead prediction pdf 
\begin{equation}
\label{eq:prior}
p(\bm x_k|\bm z_{1:k-1}) = \int p(\bm x_{k}|\bm x_{k-1}) p(\bm x_{k-1}|\bm z_{1:k-1})d\bm x_{k-1},
\end{equation}
and, when $\bm z_k$ becomes available, update the prediction to obtain the new filtering pdf 
\begin{equation}
\label{eq:posterior}
p(\bm x_k|\bm z_{1:k}) \propto p(\bm z_k|\bm x_{k})p(\bm x_k|\bm z_{1:k-1}),
\end{equation}
where $p(\bm z_k|\bm x_{k})$ is the conditional pdf of the observation $\bm z_k$ given the state $\bm x_k$ (i.e., the likelihood of $\bm x_k)$. 

If the state space model is parametrized by some random unknown parameter vector $\bm \theta$, it is possible to search for the joint posterior pdf $p(\bm x_k,\bm \theta|\bm z_{1:k})$, which admits a similar decomposition $p(\bm x_k,\bm \theta|\bm z_{1:k}) \propto p(\bm z_k|\bm x_{k},\bm \theta)p(\bm x_k,\bm \theta|\bm z_{1:k-1})$. 
The above pdfs can only be computed exactly in linear systems. Orbital dynamics, however, are a highly nonlinear system. In Section \ref{sec:methodology}, we explore several recursive MC algorithms which aim at approximating these posterior distributions numerically. 
\section{Methodology}\label{sec:methodology}
In this section, we briefly describe the principles of the simplest type of PF, the bootstrap particle filter (BPF), to illustrate the backbone of the sample-based algorithms proposed in this work, which can be used to track a spacecraft's state, as well as simultaneously estimating any unknown parameters. The Monte Carlo filters that we investigate include the ensemble Kalman filter (EnKF), which has been used in RSO tracking in \citep{Gamper}, the ensemble particle filter (EnPF), which is first introduced in this paper, and the nested hybrid particle filter (NHF) (\citep{PerezVieites2021}) algorithm, which has not been studied in problems related to astrodynamics to the best of our knowledge. We derive extensions of the EnKF and EnPF, herein denoted EnKFup and EnPFup, that enable the estimation of unknown parameters in the stochastic parametrization of Eq. \eqref{eq:sde}.

\subsection{Bootstrap particle filter}\label{sec:bpf}
The underlying principle behind PFs is Monte Carlo integration. Let $X$ be an r.v. with pdf $p(x)$. If one wishes to estimate $f(X)$ for some test function $f(\cdot)$, a natural way to proceed is to compute the expectation
\begin{equation*}
    \mathbb{E}[f(X)] = \int f({x})p(x)\text{d}x.
\end{equation*}
This expectation can be approximated by drawing $N$ samples from the pdf $p({ x})$, and then computing the sample mean
\begin{equation}
    \mathbb{E}^N[f(X)] = \frac{1}{N}\sum_{i=1}^{N}f({ x}).
\end{equation}
However, oftentimes the target distribution is not known a priori. In that case, a popular alternative is to use importance sampling (IS) schemes, where the MC samples are drawn from an importance function $q({ x})$. To account for the mismatch between $q({x})$ and $p({x})$, importance weights are assigned to these samples, given by
\begin{equation}
    \label{eq:isweights}
    w^{i} \propto \frac{p({x}^{i})}{q({x}^{i})}, \ i = 1,...,N.
\end{equation}
These weights are normalised so that $\sum_{i=1}^Nw^{i} = 1$, which implies that it is sufficient to compute $p(x^{i})$ up to a proportionality constant.
Then, the IS estimator of $\mathbb{E}[f(X)]$ is 
\begin{equation}
    \mathbb{E}^{IS,N}[f(X)] = \sum_{i=1}^{N}f({ x^{i}})w^{i}.
\end{equation}
The standard PF, often referred to as the bootstrap particle filter (BPF) or sampling importance resampling filter (SIR), is a (rather simple) sequential IS algorithm. Assume that at time $t'_{k-1}$, samples $x_{k-1}^{i}$ and weights $w_{k-1}^{i}$ have been computed. From Eq. \eqref{eq:prior}, we see that the predictive pdf $p(x_k|z_{1:k-1})$ is an integral w.r.t the pdf  $p(x_{k-1}|z_{1:k-1})$ and, therefore, it can be approximated as
\begin{equation}
     p(x_k|z_{1:k-1}) \approx p^{IS,N} ({x}_k|{z}_{1:k-1} ) := \sum_{i=1}^N w_{k-1}^i p({ x}_k|{x}_{k-1}^i ).
\end{equation}
As a consequence, the filtering pdf at time $t'_k$ can itself be estimated, namely 
\begin{equation}
    p^{IS,N}({x}_k|{ z}_{1:k} )\propto p({ z}_k|{ x}_{k}) \sum_{i=1}^N w_{k-1}^i p({ x}_k|{ x}_{k-1}^i ).
\end{equation}
If the approximate predictive pdf is used as the importance function, i.e., we draw $x_k^{i} \sim p^{IS,N}(x_k|z_{1:k-1}),\ i=1,\ldots,N$, then the importance weights at time $t_k'$ become 
\begin{equation}
    w_k^{i} \propto \frac{p^{IS,N}(x_k|z_{1:k})}{p^{IS,N}(x_k|z_{1:k-1})} \propto p(x_k^{i}|z_k),\ i=1,\ldots,N.
\end{equation}

Algorithm \ref{alg:bpf} shows an implementation of the BPF for spacecraft tracking. At time $t'_0$, $N$ particles are drawn from the prior pdf, $\bm x_0^{i}\sim p(\bm x_0),\ i=1,\ldots,N$. Then, at $t_k'$, each particle $\bm x_k^{i},\ i=1,\ldots,N$ is propagated forwards from time $t'_{k-1}$ to $t'_k$, by sampling $\tilde {\bm x}_k^{i}\sim p(\bm x_k|\tilde{\bm x}_{k-1}^{i}),\ i=1,\ldots,N$.
These predictive samples are passed through the observation function $\mathcal{M}(\cdot)$ to obtain predicted measurements ${\bm y}_k^{i}$, for $i = 1,...,N$. This enables the weights to be computed more easily as a function of the likelihoods, which are a direct proxy for how well the predicted observations match the actual observations, i.e., $w_k^{i} \propto p({\bm z}_k|{\bm x}_k^{i})\propto e^{-\frac{1}{2}(\bm z_k-\bm y_k^{i})^\top\Omega_k^{-1}(\bm z_k-\bm y_k^{i})}$. 
Resampling with replacement by using the weights of the particles follows. Particles with higher weights are randomly replicated, whilst those with lower weights are randomly discarded.  
Note that since the output of the algorithm at time $t'_k$ is the collection of weighted particles $\{\bm x_k^{i},w_k^{i}\}_{i=1}^N$ that enable the approximation of integrals w.r.t the filtering density $p(\bm x_k|\bm z_{1:k})$, it is straightforward to compute any kind of estimators, such as the posterior mean estimator, or the covariance estimator. 

The BPF represents the simplest of PFs. It often suffers from a weight degeneracy problem, where the importance weight tends to concentrate on a single particle \citep{Snyder08}. This problem can be tackled by using a large number of samples, but this is costly to run, as the computational cost of a PF is $\mathcal{O}(N)$. An alternative solution is to devise sophisticated extensions to the algorithm to bypass these drawbacks. For the time being, let it serve as motivation for the implementation of the following algorithm, the EnKF, which avoids the computation of weights altogether.

\begin{algorithm}
\caption{Bootstrap Particle Filter (BPF)}\label{alg:bpf}
\textbf{Inputs:} 
\begin{algorithmic}
\item - $N$ iid samples ${\{\bm x}_0^i \},\ i=1,\ldots,N$, from the prior pdf $p({\bm x}_0)$ at time $t_0$. 
\item - $M$ observation timestamps.
\item - A set of observations ${\bm {z}}_k$ and their noise covariance matrices $\Omega_{k}, \ k=1,...M$.
\end{algorithmic}

\textbf{Outputs:} 
\begin{algorithmic}
\item - Weighted particle sets $\{\bm {\tilde x}_k^{i},w_k^{i}\}_{i=1}^N,\ k=1,\ldots,M$.
\end{algorithmic}
\setlength\itemsep{0em}

\textbf{Procedure:} for each observation epoch $t'_k$
\vspace{0.3cm}

\begin{center} \textbf{\large Prediction} \end{center}
\begin{enumerate}
\setlength\itemsep{0em}
\item Propagate samples stochastically from time $t'_{k-1}$ to $t'_k$, using numerical scheme and dynamical model of choice, obtaining 
\newline ${\bm {\tilde x}}_k^{i} \sim p({\bm x}_k|{\bm { x}}_{k-1}^{i}), \ i = 1,...,N$.
\vspace{0.3cm}
\begin{center} \textbf{\large Update}\end{center}
\setlength\itemsep{0em}
\vspace{0.1cm}
\item Compute predicted measurements ${\bm y}_k^{i}=\mathcal{M}_k({\bm {\tilde x}}_k^{i})$ and evaluate likelihoods $L_k^{i}\sim p({\bm z}_k|{\bm {\tilde x}}_k^{i}), \ i = 1,\ldots,N.$
\vspace{0.1cm}
\item \text{Compute normalized importance weights:}
\newline $w_k^{i} \propto L_k^{i}, \ i = 1,...,N$.
\vspace{0.1cm}
\item Resample the weighted set $\{\bm{\tilde x}_k^{i},w_k^{i}\}_{i=1}^N$ $N$ times with replacement to generate new particles ${\bm x}_k^{i}, \ i = 1,...,N$.
\end{enumerate}
\end{algorithm}

Note that this version of the algorithm does not allow for parameter estimation, and is included for illustration purposes. Note, also, that the BPF is not explicitly used in Sections \ref{sec:results1} and \ref{sec:results2}.

\subsection{Ensemble Kalman filter}
The ensemble Kalman filter (EnKF) is a recursive Monte Carlo filter that replaces the computation of weights of the BPF by a particle update using an empirical Kalman gain matrix. The method has proved robust in many applications but it enjoys limited convergence guarantees compared to the PF \citep{Bishopp}.

Algorithm \ref{alg:EnKFup} follows \citep{Perez-vieites_enkf}, except that it is adapted to simultaneously track the state and a set of unknown parameters. We define the extended state vector as
\begin{equation*}
    \bm \chi_k = \left[ \begin{matrix}{\bm x}_k \\ {{\bm\theta}}\end{matrix} \right],
\end{equation*}
which includes the 6-dimensional state $\bm x_k$ and the parameter vector $\bm \theta$, increasing the dimension $d$ of the state to $D$.  Vector $\bm \theta$ contains, at least, the parameters of the diffusion coefficient $\sigma_W(\bm x)$, but it may also include dynamical parameters such as the ballistic coefficient or the SRP coefficient. 
The update stage, in contrast to the PF, uses an MC estimate of the Kalman gain instead of particle weights and resampling, meaning that all samples are equally weighted. In order to use the observation to refine the state, the Kalman gain $K_G^N$ weighs the ratio of how much the filter is to ``trust'' either the $k^{\text{th}}$ measurement or the $k^{\text{th}}$ propagated state. In this framework, a collection of $N$ samples is termed an `ensemble', which contains information about the empirical mean and covariance. These ensemble particles are then updated with a variation of the classical Kalman update equations, given by 
\begin{equation}
    {\bm {\chi}}_k^{i} = {\bm {\tilde \chi}}_k^{i} + K_G^N({\bm z}_k - {\bm y}_k^{i} + \bm s_k^{i}),\ \bm s_k^{i} \sim \mathcal{N}(0,\Omega_{k}), \ i = 1,...,N,
\end{equation}
where $\{{\bm {\tilde \chi}}_k\}_{i=1}^{N}$ are the ensemble samples before update, ${\bm z}_k - {\bm y}_k^{i}$ is the observation residual and $\bm s_k^{i}$ is the observation noise with covariance matrix $\Omega_{k}$. 
Having updated the ensemble, the estimator of the extended state is simply the mean of all particles, given by
\begin{equation}
    {\bm {\bar \chi}}_k^N = \frac{1}{N}\sum_{i=1}^N {\bm {\chi}}_k^{i},
\end{equation}
and an empirical covariance can be computed as 
\begin{equation}
    \bm C_k^N = \frac{1}{N}\sum^N_{i=1}(\bm {\chi}_k^{i}-\bar{\bm \chi}_k^N)(\bm {\chi}_k^{i}-\bar{\bm \chi}_k^N)^\top .
\end{equation}
Then, the state estimate ${\bm {\bar x}}_k^N$ and the parameter estimate ${\bm {\bar \theta}}_k^N$ can simply be extracted from ${\bm {\bar \chi}}_k^N$.
These updated ensembles are then recursively propagated to the next time step.

\begin{algorithm}
\caption{Ensemble Kalman filter with unknown parameters (EnKFup)}\label{alg:EnKFup}
\textbf{Inputs:} 
\begin{algorithmic}
\item - $N$ iid samples $\bm \chi_0^{i} ={\{\bm x}_0^i,{\bm \theta}_0^i \}, i=1,\ldots,N$, from the prior pdf $p({\bm x}_0,{\bm \theta_0})$ at time $t_0$. 
\item - $M$ observation timestamps, $\{t'_k\}^M_{k=1}$.
\item - A set of observations ${\bm {z}}_k$ and their noise covariance matrices $\Omega_{k}, k=1,...M$.
\end{algorithmic}

\textbf{Outputs:} 
\begin{algorithmic}
\item - A collection of equally-weighted samples $\{\bm \chi^{i}_k\}_{i=1}^N$ at each time $t'_k,\ k=1,\ldots,M$.
\end{algorithmic}
\setlength\itemsep{0em}

\textbf{Procedure:} for each observation epoch $t'_k$
\vspace{0.3cm}
\begin{center} \textbf{\large Prediction} \end{center}
\begin{enumerate}
    \item Propagate samples from time $t'_{k-1}$ to $t'_k$, using the numerical scheme of choice to obtain $\tilde{\bm \theta}_k^{i}=\bm \theta_{k-1}^{i}$ and ${\bm {\tilde x}}_k^{i} \sim p({\bm x}_k|{\bm x}^{i}_{k-1},\bm{\tilde \theta^{i}_k}), \ i = 1,...,N$. Let $\bm {\tilde \chi}_k^{i} ={\{\bm {\tilde x}}_k^i,{\bm {\tilde \theta}}_k^i \}$.
    \vspace{0.3cm}
\begin{center} \textbf{\large Update} \end{center}
    \item Transform the samples $\{{\bm {\tilde x}}_k^{i}\}^N_{i=1}$ through the measurement function to obtain predicted observations 
        \newline ${\bm y}_k^{i} = \mathcal{M}_k({\bm {\tilde x}}_k^{i}),\  i = 1,...,N$.
    \vspace{0.1cm}
    \item Compute mean and covariance in the observation space:
        \newline ${\bm {\hat y}}_k^{N} = \frac{1}{N}\sum_{i=1}^N {\bm y}_k^{i} \ \text{and} \ 
        C_{y,k}^N = \Omega_{k} + \frac{1}{N-1} \sum_{i-1}^N ({\bm y}_k^{i} - {\bm {\hat y}}_k^{N})({\bm y}_k^{i} - {\bm {\hat y}}_k^{N})^\top$
    \vspace{0.1cm}
    \item Compute the ensemble mean and cross-covariance matrix: 
        \newline ${\bm {\hat \chi}}_k^{N} = \frac{1}{N}\sum_{i=1}^N {\bm {\tilde \chi}}_k^{i} \ \text{and} \ 
        C_{\chi y,k}^N = \frac{1}{N-1} \sum_{i-1}^N ({\bm {\tilde \chi}}_k^{i} - {\bm {\hat \chi}}_k^{N})({\bm y}_k^{i} - {\bm {\hat y}}_k^{N})^\top$.
    \vspace{0.1cm}
    \item Compute the Kalman gain $K_G^N = C_{\chi y,k}^N (C_{y,k}^N)^{-1}$.
    \vspace{0.1cm}
    \item Update the ensemble samples
        \newline ${\bm {\chi}}_k^{i} = {\bm {\tilde \chi}}_k^{i} + K_G^N({\bm z}_k - {\bm y}_k^{i} + \bm s_k^{i}),\ \bm s_k^{i} \sim \mathcal{N}(0,\Omega_{k}), \ i = 1,...,N$.
    \vspace{0.1cm}
    
\end{enumerate}
\end{algorithm}

\subsection{Ensemble particle filter}
In order to exploit the robustness of the EnKF scheme, next, we propose a novel method that combines Kalman updates of the MC samples with the computation of importance weights, aimed at improving the accuracy of the filter. This new algorithm is termed ensemble particle filter (EnPF) as it is a hybrid between the EnKF and the BPF. Algorithm \ref{alg:EnPFup} outlines the method when it is adapted to track the state and estimate unknown parameters, similar to the EnKFup. This variation is hereby denoted EnPFup.
The prediction stage follows that of an EnKFup, which propagates an ensemble through the dynamical model. The update stage, however, now consists of two parts:
\begin{itemize}
    \item The computation of an MC estimate of the Kalman gain, in order to correct the particles based on the observation residuals, exactly as in step 2 of the EnKFup.
    \item The computation of weights (similar to the BPF) associated with these particles, which we now denote ${\bm {\check\chi}}_k^{i}$, as well as a resampling step based on these weights. In other words, by assuming that the prior is given by $\{{\bm {\check\chi}}_k\}_{i=1}^N$, weights of the form $w_k^{i} \propto w_{k-1}^{i}p(\bm z_k|\bm { \check x}_k^{i})$ are computed by calculating likelihoods, which allocate higher weights to samples that align more closely with the actual observation.
\end{itemize}   
In Algorithm \ref{alg:EnPFup}, resampling steps are taken adaptively, depending on the effective sample size (ESS), given by ESS = $\frac{1}{N\sum_{i=1}^N(w_k^{i})^2}$. The ESS is an approximate measure of sample diversity \citep{Elvira}. If the ESS falls below a given threshold $\varphi$, the particles are resampled and weights are reset.

The posterior estimate of the extended state is computed as the weighted average of all particles before resampling, namely
\begin{equation}
    \bar{{\bm {\chi}}}_k^N = \sum_{i=1}^N w_k^{i}{\bm {\check\chi}}_k^{i}.
\end{equation}
Then, the state estimate ${\bm {\bar x}}_k^N$ and the parameter estimate ${\bm {\bar \theta}}_k^N$ can simply be extracted from ${\bm {\bar \chi}}_k^N$.

\begin{algorithm}[H]
\caption{Ensemble particle filter with unknown parameters (EnPFup)}\label{alg:EnPFup}
\textbf{Inputs:} 
\begin{algorithmic}
\item - $N$ iid samples $\bm \chi_0^{i} ={\{\bm x}_0^i,{\bm \theta}_0^i \}, i=1,\ldots,N$, from the prior pdf $p({\bm x}_0,{\bm \theta_0})$ at time $t_0$. 
\item - $M$ observation timestamps, $\{t'_k\}^M_{k=1}$
\item - A set of observations ${\bm {z}}_k$ and their noise covariance matrices $\Omega_{k}, \ k=1,...M$.
\item - Resampling threshold $\varphi$.
\end{algorithmic}

\textbf{Outputs:} 
\begin{algorithmic}
\item - A collection of weighted samples $\{\bm \chi^{i}_k,w_k^{i}\}_{i=1}^{N}$, at each time $t'_k,\ k=1,\ldots,M$.
\end{algorithmic}
\setlength\itemsep{0em}

\textbf{Procedure:} for each observation epoch $t_k'$
\vspace{0.3cm}
\begin{center} \textbf{\large Prediction} \end{center}
\begin{enumerate}
\item Propagate samples from time $t'_{k-1}$ to $t'_k$, using the numerical scheme of choice to obtain $\tilde{\bm \theta}_k^{i}=\bm \theta_{k-1}^{i}$ and ${\bm {\tilde x}}_k^{i} \sim p({\bm x}_k|{\bm x}^{i}_{k-1},\bm{\tilde \theta}^{i}_k), \ i = 1,...,N$. Let $\bm {\tilde \chi}_k^{i} ={\{\bm {\tilde x}}_k^i,{\bm {\tilde \theta}}_k^i \}$.
\vspace{0.3cm}
\begin{center} \textbf{\large Update} \end{center}
    \item Transform the samples $\{{\bm {\tilde x}}_k^{i}\}^N_{i=1}$ through the measurement function to obtain predicted observations 
        \newline ${\bm y}_k^{i} = \mathcal{M}_k({\bm {\tilde x}}_k^{i}),\  i = 1,...,N$.
    \vspace{0.1cm}
    \item Compute mean and covariance in the observation space:
        \newline ${\bm {\hat y}}_k^{N} = \frac{1}{N}\sum_{i=1}^N {\bm y}_k^{i} \ \text{and} \ 
        C_{y,k}^N = \Omega_{k} + \frac{1}{N-1} \sum_{i-1}^N ({\bm y}_k^{i} - {\bm {\hat y}}_k^{N})({\bm y}_k^{i} - {\bm {\hat y}}_k^{N})^\top$
    \vspace{0.1cm}
    \item Compute the ensemble mean and cross-covariance matrix: 
        \newline ${\bm {\hat \chi}}_k^{N} = \frac{1}{N}\sum_{i=1}^N {\bm {\tilde \chi}}_k^{i} \ \text{and} \ 
        C_{\chi y,k}^N = \frac{1}{N-1} \sum_{i-1}^N ({\bm {\tilde \chi}}_k^{i} - {\bm {\hat \chi}}_k^{N})({\bm y}_k^{i} - {\bm {\hat y}}_k^{N})^\top$.
    \vspace{0.1cm}
    \item Compute the Kalman gain $K_G^N = C_{\chi y,k}^N (C_{y,k}^N)^{-1}$.
    \vspace{0.1cm}
    \item Update the ensemble samples
        \newline ${\bm {\check\chi}}_k^{i} = {\bm {\tilde\chi}}_k^{i} + K_G^N({\bm z}_k - {\bm y}_k^{i} + \bm s_k^{i}),\ \bm s_k^{i} \sim \mathcal{N}(0,\Omega_{k}), \ i = 1,...,N$.
    \vspace{0.1cm}
    \item Compute normalized importance weights in the form 
        \newline $w_k^{i} \propto w_{k-1}^{i} p({\bm z}_k|{\bm {\check\chi}}_k^{i}), \ i = 1,...,N$.
    \vspace{0.1cm}
    \item Compute the normalized ESS,
    \newline $\text{NESS}_k = \frac{1}{N\sum_{i=1}^N(w_k^{i})^2}$.
    \vspace{0.3cm}
    \item\textbf{If} $\text{NESS}_k < \varphi$, where $\varphi$ is the resampling threshold:
    \newline Resample $N$ times with replacement to generate new particles 
        ${\bm {\chi}}_k^{i}, \ i = 1,...,N$ and set $w_k^{i} = \frac{1}{N}, \ i = 1,...,N$. Otherwise, set ${\bm {\chi}}_k^{i} = {\bm{ \check\chi}}_k^{i},\ i=1,\ldots,N$.
    
\end{enumerate}
\end{algorithm}
Note that steps 1 to 6 in Algorithm \ref{alg:EnPFup} are exactly the same as Algorithm \ref{alg:EnKFup}.
It is also important to remark that this is a ``greedy" algorithm where each observation ${\bm z}_k$ is processed twice: once during the Kalman update, and again for the calculation of likelihoods that determine the weights and the resampling step. The algorithm proposed here, therefore, lacks the theoretical guarantees of convergence of conventional PFs, as the weights do not follow the principle of Eq. \eqref{eq:isweights}.

\subsection{Nested hybrid filter}
To tackle the simultaneous tracking of the states and uncertain parameters, we implement a variation of the nested particle filter (NPF) \citep{crisan1}, \citep{crisan2},  which recursively computes the posterior distribution of the parameters, $p({\bm \theta_k}|{\bm z}_{1:k} ),\ k=1,\ldots,M$.
While the EnPFup is a ``greedy" algorithm with no proven theoretical guarantees of convergence as of yet, we introduce a nested hybrid filter (NHF) for which a convergence analysis is available in \citep{PerezVieites2021}. 

The objective of the NHF is to approximate $p(\bm\theta|\bm z_{1:k})$ by employing a variation of SIR. Using Bayes, we have that 
\beq
    p(\bm\theta|\bm z_{1:k}) \propto p({\bm z}_k|\bm z_{1:k-1},\bm\theta)p(\bm\theta|\bm z_{1:k-1}).
\eeq
Therefore, if we were able to sample from $p(\bm\theta|\bm z_{1:k-1})$ and to evaluate $p(\bm z_k|\bm z_{1:k-1},\bm\theta)$, then we would be able to design a sequential importance sampler to approximate the posterior pdf, as per the methodology introduced Section \ref{sec:bpf}. Specifically, we would
\begin{enumerate}
\item sample $\bm\theta_k^{i}\sim p(\bm\theta|\bm z_{1:k-1})$,
\item compute weights $w_k^{i} \sim p(\bm z_k|\bm z_{1:k-1},\bm\theta_k^{i})$,
\item resample if necessary.
\end{enumerate}
Unfortunately, neither step 1. nor step 2. can be performed exactly. However, they can be approximated. 

Step 1. is approximated by performing a ``jittering" procedure, which consists of applying a controlled perturbation to the particles $\{{\bm \theta}_{k-1}^i\}_{i=1}^{N_1}$, resulting in $\{{\bm {\tilde\theta}}_{k}^i\}_{i=1}^{N_1}$. This perturbation is assumed to be Gaussian. It may be small and applied to many samples or it may be a relatively large perturbation applied to only a fraction of samples. See \citep{crisan1} for a detailed description and theoretical justification of the jittering step.

Step 2. is approximated by running a separate filter associated with each $\bm{\tilde\theta}_k^{i}$. Therefore, the algorithm can be visualized as a bank of $N_1$ filters (which may be PFs or Gaussian filters), one for each parameter $\bm{\tilde\theta}_k^{i}, \ i=1,\ldots,N_1$. In this implementation, the EnKF is used:
at each time $t'_k$, for each parameter sample ${\bm {\tilde\theta}}_{k}^i$, a collection of $N_2$ samples drawn from the state prior, is propagated from $t'_{k-1}$ to $t_k'$, obtaining ${\bm {\tilde x}}_k^{i,j} \sim p({\bm x}_k|{\bm x}_{k-1},\bm{\tilde \theta}_k^i),\ i = 1,...,N_1 \ \text{and}\  j = 1,...,N_2$. Then, the EnKF update equations are employed to ``move" the samples in the direction of the observation, resulting in $N_1$ ensembles of $N_2$ particles $\{\{ \bm {\check x_k}^{i,j}\}_{j=1}^{N_2}\}_{i=1}^{N_1}$.

The primary layer (i.e., the parameter layer) likelihood $p({\bm z}_k |{\bm z}_{1:k-1},{\bm {\tilde \theta}}_k^i )$ is then numerically approximated by the mean of the secondary layer (i.e., the state layer) likelihoods, as $\lambda_k^{i} = \frac{1}{N_2}\sum_{j=1}^{N_2} p({\bm z}_k|{\bm {\check x}}_k^{i,j}),\ i = 1,\ldots,N_1$, thus obtaining $N_1$ likelihoods.

All that remains is to evaluate the ESS to determine whether to resample primary layer particles (and their associated state particles), in the same way as for the EnPFup.
The NHF is featured in Algorithm \ref{alg:nhpf}.




\begin{algorithm}[H]
\caption{Nested hybrid filter (NHF)}\label{alg:nhpf}
\textbf{Inputs:} 
\begin{algorithmic}
\item - $N_1$ iid samples ${\bm \theta}_0^{i} \sim p(\bm\theta)$ and $N_2$ iid samples ${\bm x}_0^{i,j} \sim p({\bm x}_0,\bm\theta_0^i)$, for $i = 1,...,N_1$ and $j=1,\ldots,N_2$ at time $t'_0$. 
\item - Initial weights $w_0^{i} = \frac{1}{N_1}, i = 1,...,N_1$.
\item - $M$ observation timestamps, $\{t'_k\}_{k=1}^M$
\item - A set of observations ${\bm {z}}_k$ and their noise covariance matrices $\Omega_{k}, k=1,...M$.
\item - Resampling threshold $\varphi$.
\end{algorithmic}

\textbf{Outputs:} 
\begin{algorithmic}
\item A collection of weighted samples $\{\bm \chi^{i}_k,w_k^{i}\}_{i=1}^{N_1},\ k = 1,\ldots,M$, where $\bm \chi^{i}_k = \{\bm \theta_k^{i}, \bm x_k^{i,1:N_2}\}$, for $i = 1,\ldots,N_1$.
\end{algorithmic}
\setlength\itemsep{0em}
\textbf{Procedure:} for each observation epoch $t'_k$
\vspace{0.3cm}
\begin{center} \textbf{\large Prediction} \end{center}
\begin{enumerate}
    \item \textbf{Jittering}: Generate new particles in the parameter space by computing:
        \newline ${\bm{\tilde\theta}}_k^{(i)} = 
        \begin{cases} 
        \bm\theta_{k-1}^{(i)} & \text{with probability } 1 - \epsilon_{N_1}, \\
        \bm\theta_{k-1}^{(i)} + \kappa_k^{(i)} & \text{with probability } \epsilon_{N_1},
        \end{cases}$
        for $i = 1, \dots, N_1$, 
        \newline where $\epsilon_{N_1} = N_1^{-1/2}$ and $\kappa_k^{(i)}$ is an independent noise term.
        \vspace{0.1cm}
    \item Propagate samples ${\bm x}_k^{i,j}$ from time $t'_{k-1}$ to $t'_k$, using a numerical scheme of choice, obtaining ${\bm {\tilde x}}_k^{i,j} \sim p({\bm x}_k|{\bm x}^{i,j}_{k-1},\bm{\tilde\theta}_{k}^i), i = 1,...,N_1 \ \text{and}\  j = 1,...,N_2$. 
\vspace{0.3cm}
\begin{center} \textbf{\large Update} \end{center}
    \vspace{0.2cm}
 \textbf{FOR $\bm i = \bm 1,\ldots,\bm N_1$:}
    \vspace{0.1cm}
    \item  Transform the samples $\{{\bm {\tilde x}}_k^{i,j}\}^{N_2}_{j=1}$ through the measurement function to obtain predicted observations 
        \newline ${\bm y}_k^{i,j} = \mathcal{M}_k({\bm {\tilde x}}_k^{i,j}),\  j = 1,...,N_2$.
    \vspace{0.1cm}
    \item Compute mean and covariance in the observation space:
        \newline ${\bm {\hat y}}_k^{i,N_2} = \frac{1}{N_2}\sum_{j=1}^{N_2} {\bm y}_k^{i,j} \ \text{and} \ 
        C_{y,k}^{i,N_2} = \Omega_{k} + \frac{1}{N_2-1} \sum_{j-1}^{N_2} ({\bm y}_k^{i,j} - {\bm {\hat y}}_k^{i,N_2})({\bm y}_k^{i,j} - {\bm {\hat y}}_k^{i,N_2})^\top$
    \vspace{0.1cm}
    \item Compute the ensemble mean and cross-covariance matrix: 
        \newline ${\bm {\hat x}}_k^{i,N_2} = \frac{1}{N_2}\sum_{j=1}^{N_2} {\bm {\tilde x}}_k^{i,j} \ \text{and} \ 
        C_{\chi y,k}^{i,N_2} = \frac{1}{N_2-1} \sum_{j=1}^{N_2} ({\bm {\tilde x}}_k^{i,j} - {\bm {\hat x}}_k^{i,N_2})({\bm y}_k^{i,j} - {\bm {\hat y}}_k^{i,N_2})^\top$.
    \vspace{0.1cm}
    \item Compute the Kalman gain $K_G^{i,N_2} = C_{\chi y,k}^{i,N_2} (C_{y,k}^{i,N_2})^{-1}$.
    \vspace{0.1cm}
    \item Update the ensemble samples
        \newline ${\bm {\check x}}_k^{i,j} = {\bm {\tilde x}}_k^{i,j} + K_G^{i,N_2}({\bm z}_k - {\bm y}_k^{i,j} + \bm s_k^{i,j}),\ \bm s_k^{i,j} \sim \mathcal{N}(0,\Omega_{k}), \ j = 1,...,N_2$.
    \vspace{0.1cm}
    \item Compute primary (parameter) layer weights as \newline $w_k^{i} \propto w_{k-1}^{i}\lambda_k^{i}$, where $\lambda_k^{i} = \frac{1}{N_2}\sum_{j=1}^{N_2} p({\bm z}_k|{\bm {\check x}}_k^{i,j})$
    \vspace{0.1cm}
\newline \textbf{END FOR}
\vspace{0.2cm}
    \item Let $\bm {\check\chi}^{i}_k = [\bm {\theta}_k^{i}, \bm {\check x}_k^{i,1:N_2}]^\top$.
    \vspace{0.1cm}
    \item Compute the normalized ESS,
    \newline $\text{NESS}_k = \frac{1}{N_1\sum_{i=1}^{N_1}(w_k^{i})^2}$.
    \vspace{0.3cm}
    \item\textbf{If} $\text{NESS}_k < \varphi$,
    \newline Resample $N_1$ times with replacement to generate new particles $\{\bm \chi_k^{i}\}_{i=1}^{N_1}$, where $\bm \chi_k^{i} = \{\bm \theta_k^{i}, \bm x_k^{i,1:N_2}\}$ and set $w_k^{i} = \frac{1}{N_1}, i = 1,...,N_1$. Otherwise, set $\bm \chi_k^{i} = \bm {\check \chi}_k^{i}$.
        
\end{enumerate}
\end{algorithm}

\section{Test case}\label{sec:test_case}
\subsection{Initial conditions and simulation setup}\label{sec:test_case1}
The test case used for the validation of the above algorithms is a LEO scenario with synthetic radar observations generated with a high-fidelity (HF) dynamical model, spanning a period of just over 9 days. The object's physical parameters are outlined in Table \ref{tab:LEOphysicalparams}. 
\begin{table}[h]
\caption{Physical properties of the LEO spacecraft} 
\normalsize
\centering 
\begin{tabular}{c c c} 
\hline\hline 
& Object 1 \\ 
\hline 
Semi-major axis (km) & 6879.602 \\
Altitude (km) & 501.466 \\
Mass (kg) & 2.678 \\ 
Area ($m^2$) & 0.0402 \\ 
Drag coefficient (-) &  2.667 \\ 
\hline 
\end{tabular}
\label{tab:LEOphysicalparams}
\end{table}

The initial state $\textbf{x}(t_0)$ is given by
\begin{equation*}
\mathbf{x}(t_0)
=
\begin{aligned}
\begin{array}{ cccccc }
  [\text{-}2815.17\text{km} & 6200.05\text{km} & \text{-}967.78\text{km} & 0.15\text{km/s} & \text{-}1.09\text{km/s} & \text{-}7.53\text{km/s}]^\top, \\ 
\end{array}
\end{aligned}
\end{equation*}
and the initial covariance $\mathbf{\Sigma}_0$ is given by
\begin{equation*}
\mathbf{\Sigma}_0 =
\begin{bmatrix}
0.05 & -0.08 & 0.02 & 0 & 0 & 0 \\
-0.08 & 0.18 & -0.03 & 0 & 0 & 0 \\
0.01 & -0.03 & 0.01 & 0 & 0 & 0 \\
0 & 0 & 0 & 1.53\times 10^{-8} & 4.81\times 10^{-9} & -5.29\times 10^{-9} \\
0 & 0 & 0 & 4.81\times 10^{-9} & 1.04\times 10^{-8} & 3.46\times 10^{-8} \\
0 & 0 & 0 & -5.29\times 10^{-9} & 3.46\times 10^{-8} & 2.42\times 10^{-7}
\end{bmatrix} \!
\begin{array}{l}
\text{km}^2 \\
\text{km}^2 \\
\text{km}^2 \\
\text{km}^2/\text{s}^{2} \\
\text{km}^2/\text{s}^{2} \\
\text{km}^2/\text{s}^{2}.
\end{array}
\end{equation*}

The parameter $\bm\theta$ introduced in Section \ref{sec:methodology} takes the form of the diffusion coefficient $\sigma_W$ of Section \ref{sec:dynmod}. Furthermore, it is assumed that the parameter is equivalent in all three directions of the velocity, i.e., 
\begin{equation*}
\sigma_{W} \equiv \sigma_{W(v_x)}=\sigma_{W(v_y)}=\sigma_{W(v_z)},
\end{equation*}
\newline and hence the problem is reduced to estimating a scalar. The initial support of $\sigma_W$ is given as $[ 10^{-10}, 10^{-6} ]$, to account for the expected scale of the truncated accelerations in the dynamical model.

A reference orbit is required to be able to generate synthetic observations. This orbit may be computed deterministically, or stochastically.
In this work, we conduct validation experiments on two different simulation setups:
\begin{itemize}
\item In the first scenario, we aim to show that the proposed methods can correctly estimate the nominal parameter used in the generation of the reference orbit. To this aim, HF stochastic propagation is used for the generation of the reference orbit, including a fixed stochastic diffusion coefficient $\sigma_W = 5 \times 10^{-8}$. Within the filters, an HF dynamical model is used in the stochastic propagation scheme.
\item In the second scenario, we test whether the proposed algorithms can yield a diffusion coefficient that accounts for a model mismatch. In this case, the reference orbit is computed deterministically (i.e., with $\sigma_W = 0$) using an HF model, while the filters are implemented with a low-fidelity (LF) stochastic model. This is done in order to evaluate the ability of the algorithms to characterize the diffusion magnitudes required to mitigate the impact of using a reduced fidelity model.
\end{itemize}

In all cases, the HF model used is the ``high precision orbit propagator library" (HPOP) in MATLAB \citep{Mahooti}, which applies the NRLMSISE-00 drag model, the GGM03C model for Earth gravitational potential of up to order and degree 10, Luni-Solar perturbations, as well as other planets' gravitational influence and relativity perturbations. The LF model used in the second setup is the HPOP library, but applying only the NRLMSISE-00 drag model and the gravitational potential of up to order and degree 5.
The computations for the proposed methods on both setups are run on a Macbook Pro with an Apple M1 processor. 

\subsection{Observations}\label{sec:obs}
The synthetic observations used for the filtering process are noisy radar measurements obtained from the trajectory generated by HF propagation of the initial conditions. The measurement set is obtained by transforming states via an observation function, implemented in MATLAB, which converts ECI coordinates into the four variables which make up typical LEO observations: the range (m), azimuth (rad), elevation (rad) and range-rate (m/s). These transformations are acquired in irregularly-spaced intervals, with the minimum observation time separation being 30 seconds, and the maximum time separation being just over 9 hours. The entire simulation time-span is 9 days, from 00:01 on the 15th of August to 10:01 on the 24th of August, 2023.
The covariance matrix of the measurement noise is fixed and given by $\Omega = \text{diag}[0.02,\ 0.007,\ 0.007,\ 0.0007]^2$.

\section{Results}
\subsection{Estimating the nominal parameter}\label{sec:results1}
In this section, we tackle the first scenario described in Section \ref{sec:test_case1}. In particular, we assess whether the proposed filters can estimate the noise parameter, $\sigma_W$, with sufficient accuracy. This is done by comparing the estimate provided by each filter with the ground-truth parameter $\sigma_W$ used to generate the synthetic data. 
Performance assessment is carried out in terms of:
\begin{itemize}
\item The root-mean square errors (RMSE) in position (m) and velocity (m/s) of the algorithms described in Section \ref{sec:methodology}. These are calculated for both predicted or estimated states with respect to the reference orbit, at each observation timestamp.
\item The computational run-time of each algorithm simulation.
\end{itemize}
In order to provide a benchmark, the EnKF and EnPF can be run with models in which there are no unknown parameters. In addition, a nonlinear batch least squares (NLBLS) is run a number of times with samples drawn from the initial distribution.
The above metrics are helpful when determining the correct number of samples to be used in a trade-off between accuracy and speed. As long as RMSE values show no extreme differences, and none of the filters diverge, computational speed takes precedence for the assessment of performance.
The number of samples for each of the algorithms, following these criteria are $N_{\text{EnKFup}} = 450$, $N_{\text{EnPFup}} = 350$ and $N_{\text{NHF}} = 50\times50$ (i.e., $N_1=N_2=50$).
In the implementation of the NHF, the jittering step is performed by setting $\epsilon_{N_1} = N_1^{-\frac{1}{2}}$, and taking $\kappa_k^{(i)}$ to be Gaussian noise with zero mean and covariance set to $0.25$.

\subsubsection{Parameter estimation}\label{sec:param_est}
\begin{figure}[H]
    \centering
    \includegraphics[scale=0.23]{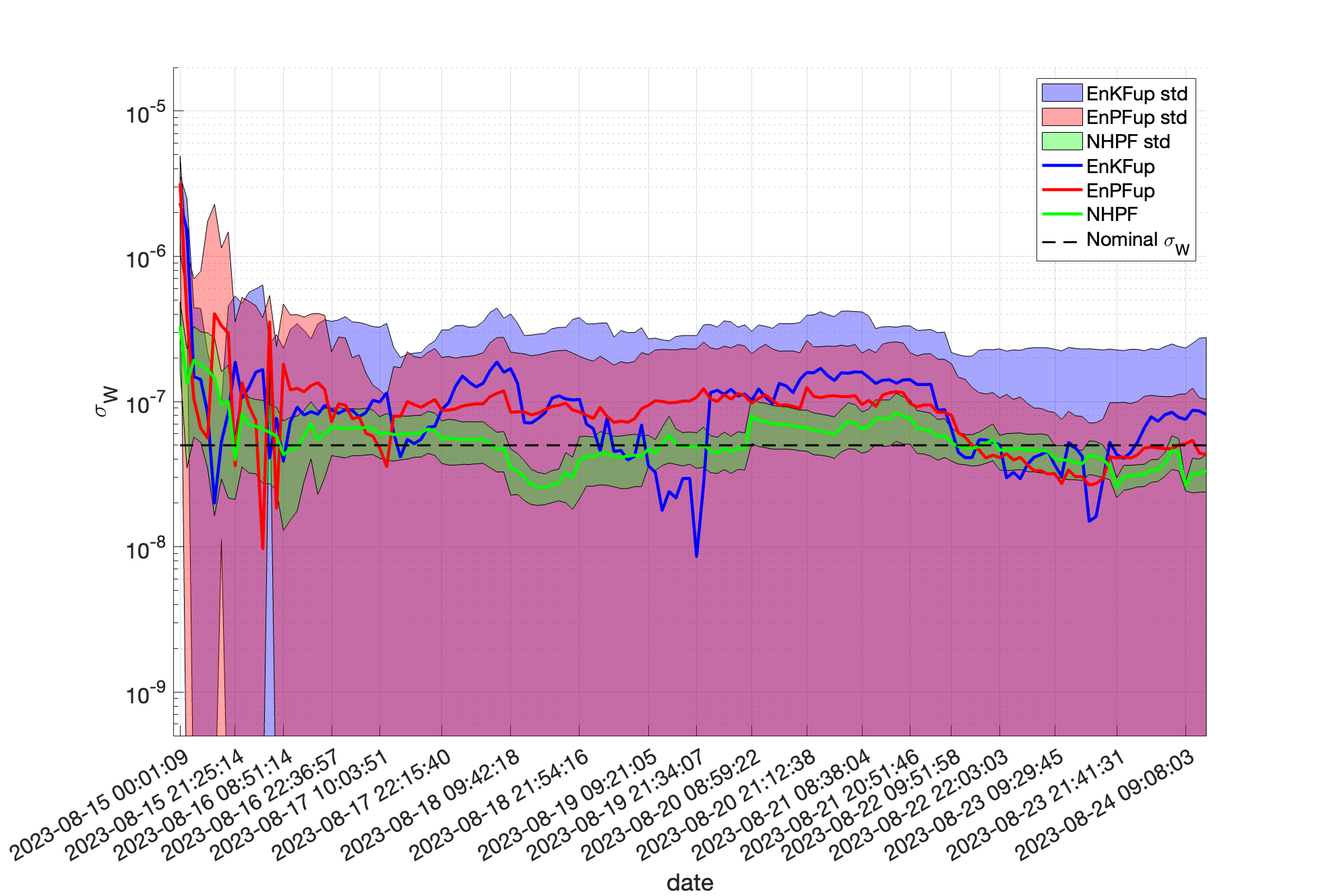}
    \caption{Parameter estimates shown in solid lines for the EnKFup (blue), the EnPFup (red) and the NHF (green), compared with the ground truth values (dotted line). In addition, shaded areas representing the standard deviations are shown with the corresponding colours of each algorithm.}
    \label{fig:algorithms_sigma_1}
\end{figure}

The estimates of $\sigma_W$ for each filter are shown in Figure \ref{fig:algorithms_sigma_1}, along with the corresponding standard deviations computed over 30 independent simulations. The black dotted line is the ground truth, given by $\sigma_W = 5 \times 10^{-8}$ in units of acceleration, i.e., $\text{km}^2/\text{s}^2$. Initial values $\sigma_{W,0}^{i},\ i = 1,\ldots,N$ are drawn from the parameter prior $p(\sigma_{W,0})$, with support $[10^{-10},\ 10^{-6}]$.

All filters adjust the order of magnitude relatively quickly. Large fluctuations can be observed in the EnKFup (blue line) with a large variance (blue shaded region), indicating a relative lack of precision compared to the other two algorithms.  
Nevertheless, even with these fluctuations, all three filters are able to converge acceptably well to the correct value (black dotted line) and manage to achieve sufficiently low error rates in position. The most precise estimator of $\sigma_W$ is the NHF, as seen by the standard deviation (green) around the estimate, as well as the estimator with the lowest bias throughout the studied time-span. However, the EnPFup achieves a reasonably accurate estimate (though it requires a longer simulation time to do so), and its overall computational cost is considerably lower than that of the NHF.

\begin{figure}[H]
    \centering
    \includegraphics[scale=0.62]{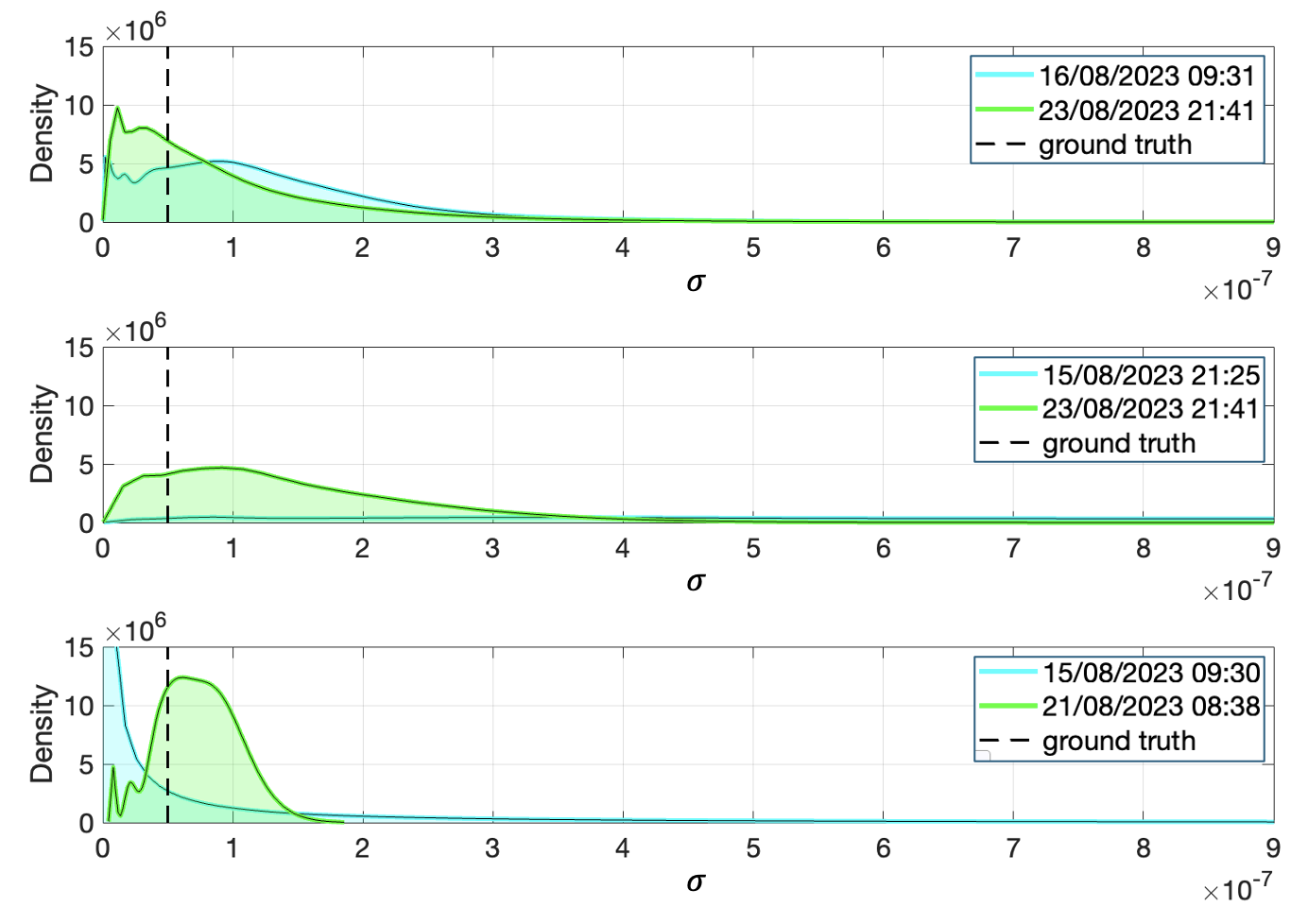}
    \caption{Estimated posterior pdf of $\sigma_W$ for the EnKFup (top), the EnPFup (middle) and the NHF (bottom). Two time-steps are shown: in cyan, a time step before convergence ($\sim 21:00$, 15th of August 2023), where a larger spread and hence worse estimate can be seen, and in green, a time step after convergence ($\sim 21:40$, 23rd of August, 2023), with more concentration of samples around the ground truth value (dotted line).}
    \label{fig:algorithms_density_sigma_1}
\end{figure}

In Figure \ref{fig:algorithms_density_sigma_1}, an estimate of the parameter posterior pdf for all three algorithms is shown at two different instants: in cyan, the particles before the algorithm has converged (corresponding to one of the initial time-steps in Figure \ref{fig:algorithms_sigma_1}), and in green, the particles at some time step after convergence near the ground-truth value. A visibly larger spread towards large values can be observed for the cyan curve, and a more concentrated area can be observed for the green curves, indicating a denser concentration of samples around the ground-truth value (dotted line). While for the EnKFup the increase in concentration of probability mass is less distinguishable between the two time steps, the EnPFup and NHF show a more pronounced difference in curve peak location and concentration.

\subsubsection{Position and velocity errors}\label{sec:pos_err1}
\begin{figure}[H]
    \centering
    \includegraphics[scale=0.24]{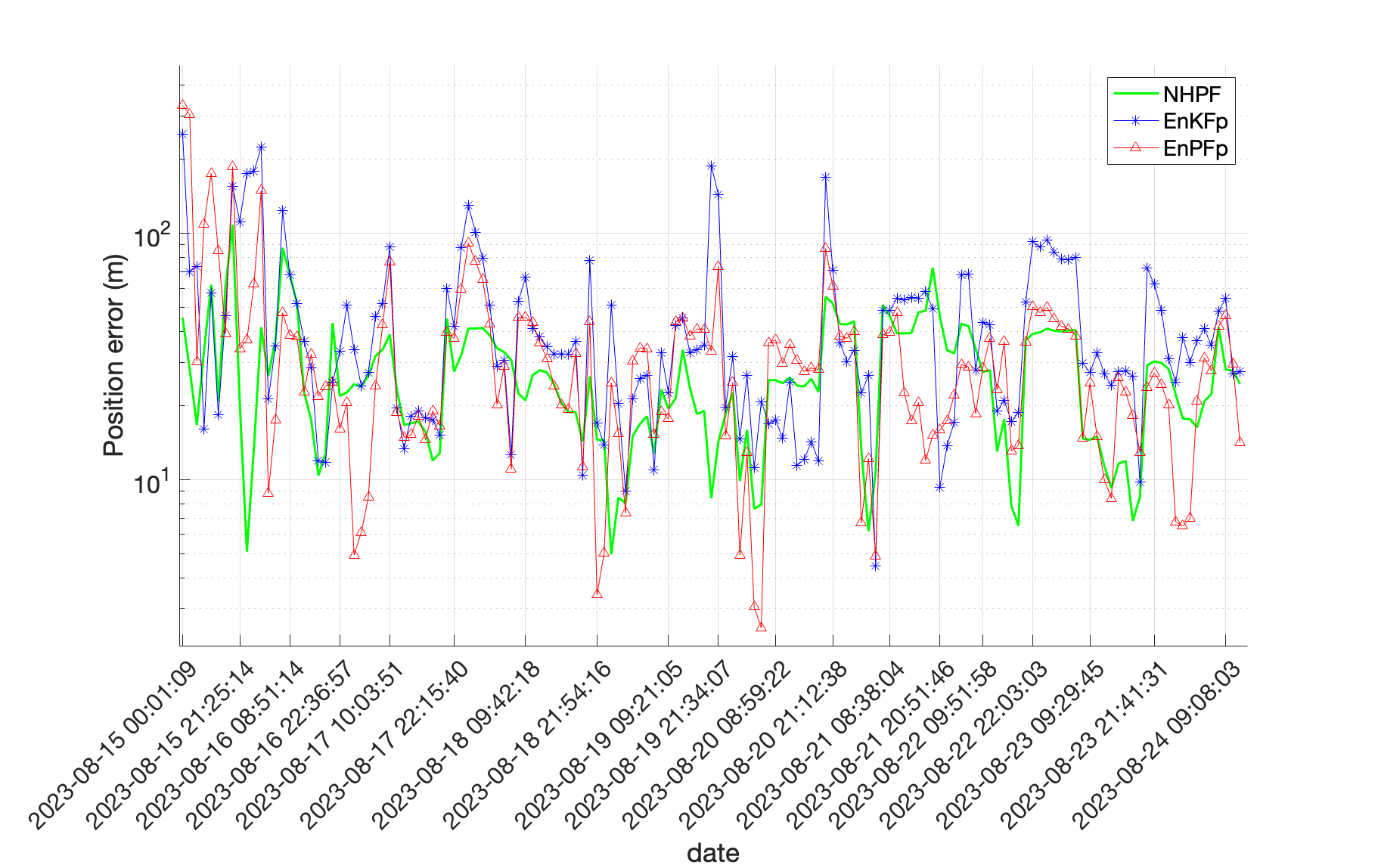}
    \caption{RMSEs in position for the three proposed algorithms which estimate the given $\sigma_W$. The solid blue line shows the EnKFup errors, the red line shows the EnPFup errors, and the green line shows the NHF errors.}
    \label{fig:all_pos_1}
\end{figure}

\begin{figure}[H]
    \centering
    \includegraphics[scale=0.25]{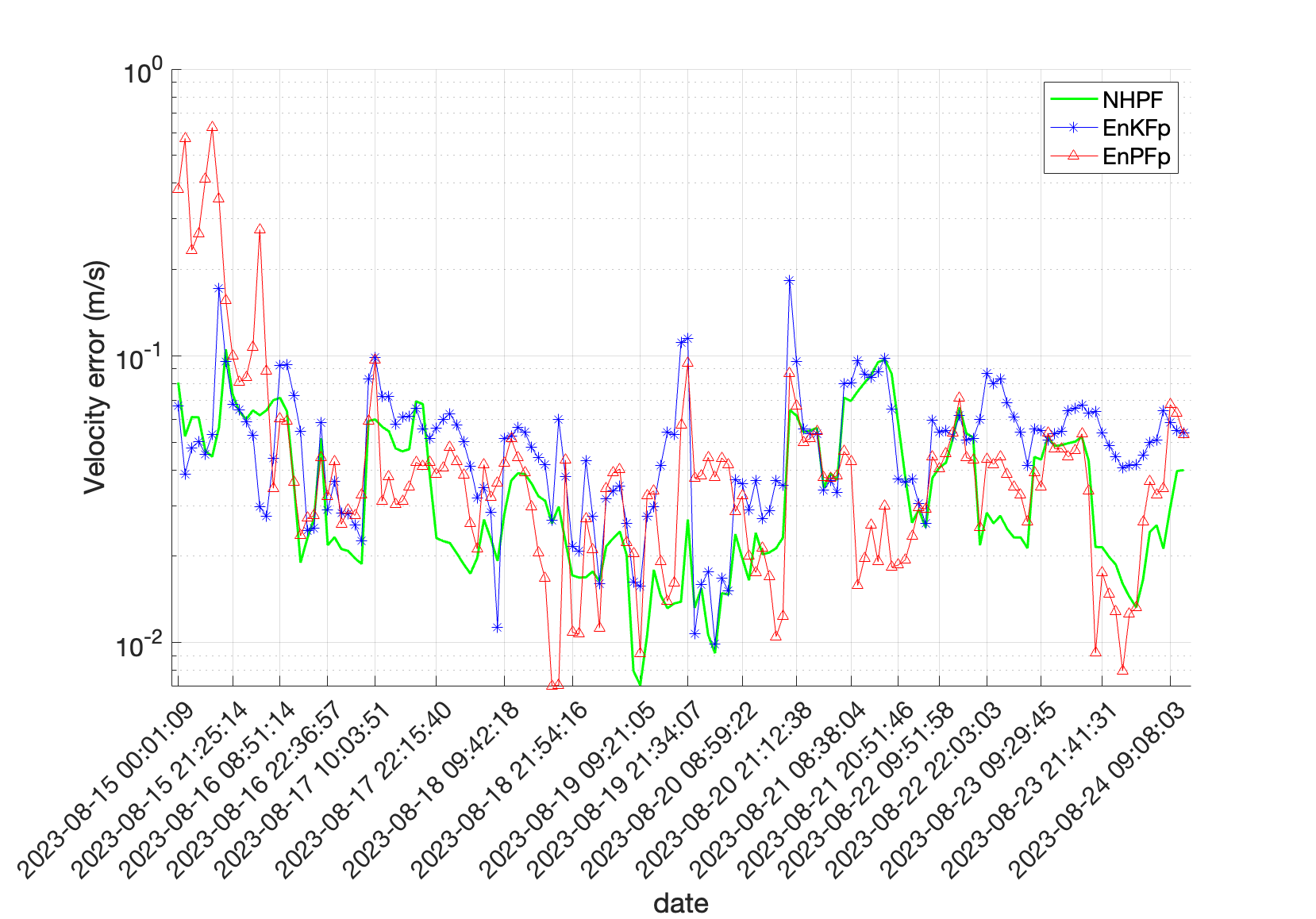}
    \caption{ RMSEs in velocity for the three proposed algorithms which estimate the given $\sigma_W$. The solid blue line shows the EnKFup errors, the red line shows the EnPFup errors, and the green line shows the NHF errors.}
    \label{fig:all_vel_1}
\end{figure}

In this subsection, the position and velocity RMSEs are shown for the entirety of the simulation period (9 days), in order to observe the trend in the accuracy of the algorithms as the parameter is adjusted in the background. 
Figure \ref{fig:all_pos_1} shows the RMSEs in position, while Figure \ref{fig:all_vel_1} shows the RMSEs in velocity of the three algorithms. The EnKFup is shown as a solid blue line with star pointers, the EnPFup as a solid red line with triangle pointers, and the NHF is shown as a thick solid green line.

For the three algorithms, the estimated errors fluctuate significantly throughout the propagation due to relatively long re-observation times. This means a much longer observation-free propagation between tracklets than between observations in the same tracklet. A slight downward tendency can be observed, both in the position and in the velocity errors, indicating not only a correct and satisfactory performance of the filter, but also an adjustment of the unknown parameter as it is being estimated in the background. 
Both the EnKFup and EnPFup begin with large errors in position (over 120m), which are in stark contrast to the final values (20-30m), but achieve errors as low as 3m throughout the simulation.

\subsubsection{Performance comparison}
The computational run-time for each algorithm is shown in Table \ref{tab:Algo_perf1} below. Note that the period of simulation is 9 days, which is a significant propagation period, and the reason why the run-times are relatively high. The results shown are the averages of 10 simulations for each algorithm. Shown as well, are the RMSEs in the form $\epsilon \pm \sigma_{\epsilon}$, where $\epsilon$ is the mean error and $\sigma_{\epsilon}$ is the standard deviation. 
The table also includes results for the ensemble Kalman filter (EnKF) and the ensemble particle filter (EnPF) with known parameter $\sigma_W$. These algorithms are implemented as in Algorithm \ref{alg:EnKFup} and Algorithm \ref{alg:EnPFup}, respectively, except that the unknown state reduces to the position and velocity of the object, that is, $\bm\chi_k = \bm x_k$.
The table includes the results for the same algorithms but assuming the nominal parameter, which simply tracks the state, and assumes the correct $\sigma_W$ value. Additionally, Table \ref{tab:Algo_perf1} also shows the errors obtained by a nonlinear batch least squares (NBLS) algorithm applied to the same initial conditions. 
\footnote{The NHF is easily paralellizable using MATLAB's parfor capabilities for parallel computing. In this case, it is done over the parameter space, so that the state space EnKFs run in parallel.}

\begin{table}[h]
\caption{Performance comparison for the three algorithms. The metrics are the root mean squared error (RMSE) in position and velocity of the estimated trajectory compared to the reference trajectory, the error standard deviations included as $\epsilon \pm \sigma_{\epsilon}$ and the mean run-time.} 
\normalsize
\begin{center}

    \hspace*{0.1cm}\begin{tabular}{l c c c}
    Algorithm & Position RMSE (m) & Velocity RMSE (m/s) & Run-time (min) \\
    \hline
    EnKFup      & $44.31 \pm 9.12$ & $0.07 \pm 0.004$ &  35.32\\
    EnKF       & $10.15 \pm 3.52$ & $0.01 \pm 0.001$&  34.50 \\
    EnPFup      & $40.19 \pm 11.13$ & $0.06 \pm 0.002$&  41.51 \\
    EnPF       & $14.29 \pm 3.52$ & $0.02 \pm 0.002$& 39.42  \\
    NHF       & $17.71 \pm 7.88$ & $0.01 \pm 0.002$ & 351.29 \\
    EnKF(50)*  & $10.77 \pm 4.79$ & $0.01 \pm 0.001$ & 12.21  \\
    \hline
    {\bf NLBLS}      & $54.38 \pm 16.81$ & $0.06 \pm 0.04$ & 12.44 \\
    \hline
    \label{tab:Algo_perf1}
    \end{tabular}
    *The EnKF(50) run involves using the same number of samples for the state, $N = 50$, as the NHF run.
\end{center}
\end{table}

The NHF is by far the most costly, showing a run-time 1 order of magnitude higher than the rest. The EnKFup and EnPFup show similar run-times, with the EnPFup attaining slightly smaller errors. Together with the results in Section \ref{sec:param_est}, this seems to indicate a better trade-off between computational cost and accuracy for this algorithm, achieving errors which are comparable to a nonlinear batch least squares filter with known parameters, an industry standard. The nominal parameter versions of the algorithms (EnKF and EnPF) achieve a lower RMSE, as expected. 

The errors in Table \ref{tab:Algo_perf1} represent the estimation RMSE of the corresponding trajectory, upon being updated by each incoming measurement. However, prediction errors (before the measurement update) can be calculated by using a single forward pass of a propagation model using the Qlaw method \citep{Qlaw}. The idea behind this is to obtain a dynamics-informed interpolation between estimated values, by calculating appropriate acceleration magnitudes in orbital elements at $t_{k-1}$ in order to drive the state towards the estimated target value at time $t_{k}$. An RMSE value can be calculated at this point between this corrected trajectory and the reference trajectory.

The results for each algorithm are shown in Table \ref{tab:alg_pred}, and are given as the average errors throughout the entire run. These are inevitably higher than the estimation errors, as the latter are calculated after the processing of observations. The relative performance of the filters is the same as described for Table \ref{tab:Algo_perf1}, with the NHF attaining the lowest errors (at the highest computational cost) and the EnPFup achieving the best trade-off between accuracy and computational cost.
\begin{table}[h]
\caption{Prediction errors for the three algorithms. The metrics are the root mean squared error (RMSE) in position and velocity and the error standard deviations included as $\epsilon \pm \sigma_{\epsilon}$.} 
\normalsize
\begin{center}

    \hspace*{0.1cm}\begin{tabular}{l c c c}
    Algorithm & Position pred. RMSE (m) & Velocity pred. RMSE (m/s) \\
    \hline
    EnKFup      & $110.21 \pm 12.47$ & $0.09 \pm 0.02$ \\
    EnPFup      & $104.34 \pm 10.11$ & $0.08 \pm 0.03$  \\
    NHF       & $102.41 \pm 11.29$ & $0.07 \pm 0.02$  \\
    \hline
    \label{tab:alg_pred}
    \end{tabular}
\end{center}
\end{table}

\subsection{Estimating the parameter with simplified dynamics}\label{sec:results2}
In this section, the performance of the proposed filters is evaluated using a simplified dynamical model. To do this, an HF reference orbit is generated deterministically. This reference orbit is also used to produce synthetic radar observations (as discussed in Section \ref{sec:obs}). However, the different filters (EnKFup, EnPFup, NHF) are built around a stochastic LF representation of the orbital dynamics. The goal is to assess whether the filtering algorithms can estimate a suitable noise parameter $\sigma_W$ that accounts for the difference between the LF model used by the algorithms and the HF model that generates the reference orbit. 
This is done by computing the RMSEs in position (m) and velocity (m/s) of all three filters. In addition, the computational run-times are assessed to determine the optimal trade-off of accuracy and speed. 
The number of samples used for the EnKFup, the EnPFup, and the NHF are $N_{\text{EnKFup}} = 450$, $N_{\text{EnPFup}} = 450$ and $N_{\text{NHF}} = 50\times50$ (i.e., $N_1=N_2=50$).

\subsubsection{Position and velocity errors}\label{sec:pos_err2}
\begin{figure}[H]
    \centering
    \includegraphics[scale=0.27]{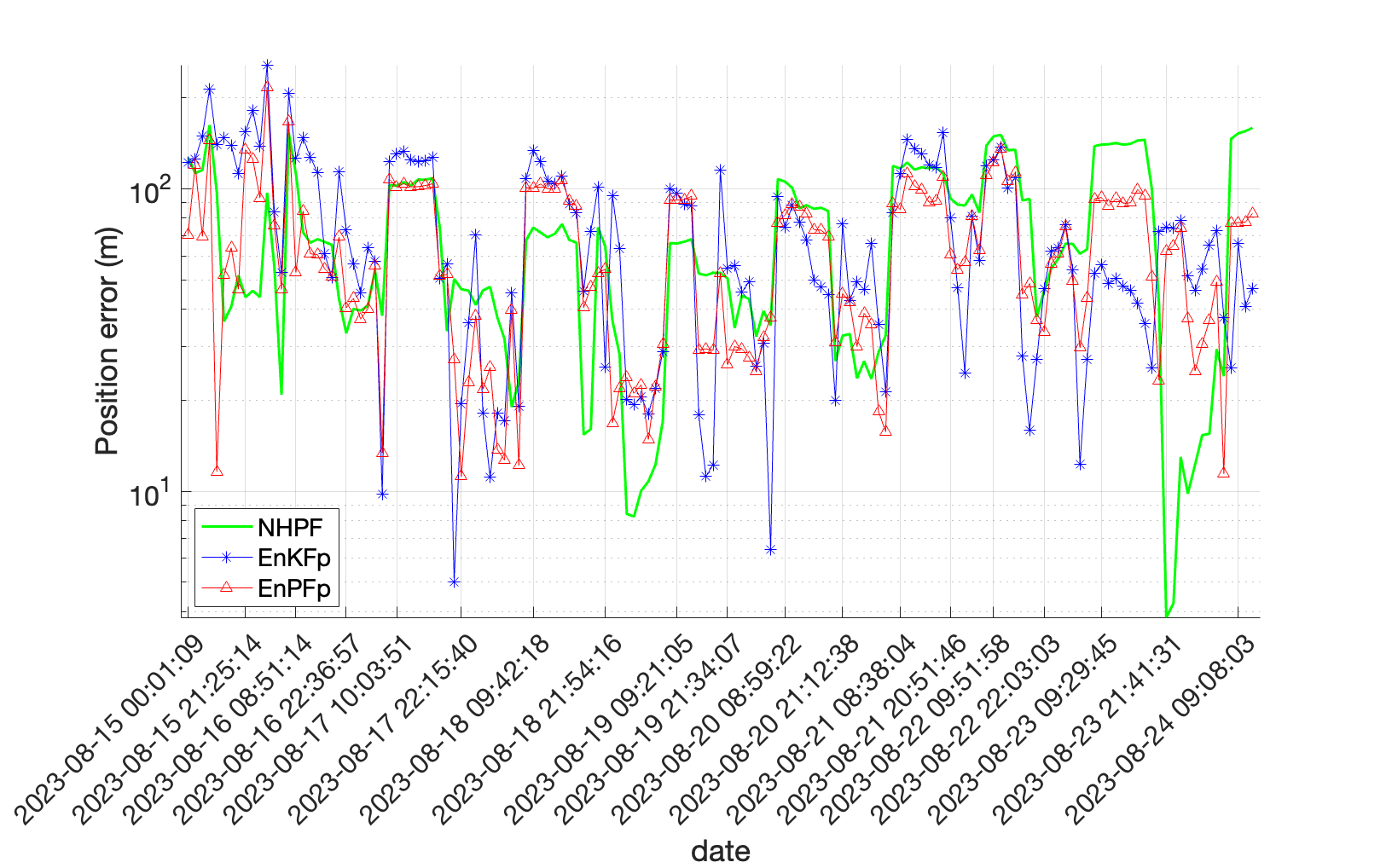}
    \caption{RMSEs in position for the three proposed algorithms which adjust the process noise magnitude, $\sigma_W$ to account for the difference between the HF model used to generate the reference orbit and the LF model used by the filters. The solid blue line shows the EnKFup errors, the red line shows the EnPFup errors, and the green line shows the NHF errors.}
    \label{fig:all_pos_2}
\end{figure}

\begin{figure}[H]
    \centering
    \includegraphics[scale=0.27]{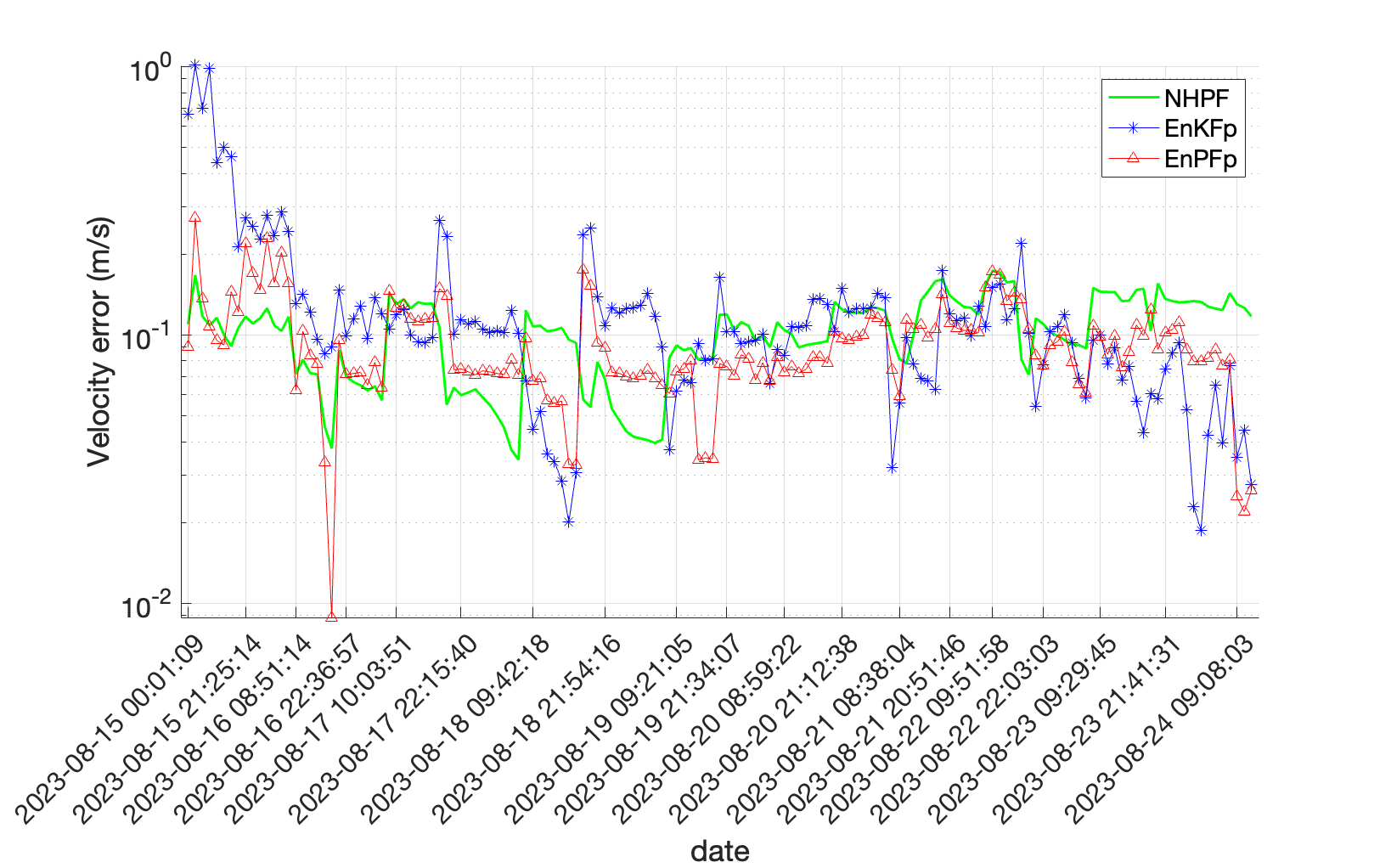}
    \caption{RMSEs in velocity for the three proposed algorithms which adjust the process noise magnitude, $\sigma_W$, to account for the difference between the HF model used to generate the reference orbit and the LF model used by the filters. The solid blue line shows the EnKFup errors, the red line shows the EnPFup errors, and the green line shows the NHF errors.}
    \label{fig:all_vel_2}
\end{figure}

In this subsection, the position and velocity RMSEs are shown for the entirety of the simulation period (9 days), to observe the trend in the accuracy of the algorithms as the parameter is simultaneously adjusted. 
Figure \ref{fig:all_pos_2} shows the RMSEs in position and Figure \ref{fig:all_vel_2} shows the RMSEs in velocity of the three algorithms.
Compared to Figures \ref{fig:all_pos_1} and \ref{fig:all_vel_1}, the estimation errors display larger fluctuations over time, due to the simultaneous adjustment of the process noise to account for the simplified model being used. 
A clear adjustment takes place almost immediately after the start of the propagation period. The filter shows to have somewhat converged after 2 days and remains at low error values for most of the simulation. These figures demonstrate the ability to use LF models and achieve sufficiently low RMSEs when using a filter which estimates $\sigma_W$, hence decreasing the computational cost associated with tracking an object online.

\subsubsection{Performance comparison}
The RMSEs and computational run-time for each algorithm are shown in Table \ref{tab:Algo_perf2}. Note that the period of simulation is 9 days, which is a significant propagation period, and the reason why the run-times are relatively high. The results presented in the table are the averages over 10 simulations of each of the algorithms, showing, in addition, the standard deviations of the position and velocity RMSEs, and the mean run-time.
Prediction errors are shown in Table \ref{tab:alg_pred2}. They are computed in the same way as for Table \ref{tab:alg_pred}.

\begin{table}[h]
\caption{Performance comparison for the three algorithms. The metrics are the RMSE in position and the RMSE in velocity of the estimated trajectory compared to the reference trajectory, the standard deviations of the obtained errors, and the run-time (in minutes).} 
\normalsize
\begin{center}
    \hspace*{0.1cm}\begin{tabular}{l c c c}
    Algorithm & Position RMSE (m) & Velocity RMSE (m/s) & Run-time (min)\\
    \hline 
    EnKFup & $70.71 \pm 13.19$ & $0.10 \pm 0.05$ & $17.50$\\ 
    EnPFup & $61.60 \pm 9.29$ & $0.09 \pm 0.04$ & $18.21$\\
    NHF & $ 69.18 \pm 7.81$ & $0.10 \pm 0.01$ & $141.36$ \\
    \hline 
    \end{tabular}
    \label{tab:Algo_perf2}
\end{center}
\end{table}

The NHF is still by far the most computationally costly of the three methods. Both the EnKFup and EnPFup show similar run-times, and therefore, given also the results in Section \ref{sec:param_est}, the EnPFup attains the best trade-off between computational efficiency and accuracy, as it achieves the lowest errors in position and a run-time one order of magnitude lower than the NHF.

\begin{table}[h]
\caption{Prediction errors for the three algorithms. The metrics are the RMSE in position and velocity. The error standard deviations are indicated as $\pm \sigma_{\epsilon}$.} 
\normalsize
\begin{center}
    \hspace*{0.1cm}\begin{tabular}{l c c c}
    Algorithm & Position pred. RMSE (m) & Velocity pred. RMSE (m/s) \\
    \hline
    EnKFup      & $169.13 \pm 21.18$ & $0.14 \pm 0.06$ \\
    EnPFup      & $159.81 \pm 13.33$ & $0.11 \pm 0.04$  \\
    NHF       & $174.19 \pm 10.50$ & $0.14 \pm 0.03$  \\
    \hline
    \label{tab:alg_pred2}
    \end{tabular}
\end{center}
\end{table}

\subsection{Estimating a 3-dimensional parameter}
In this section, the noise parameter $\sigma_W$ of the stochastic LF model consists of three different quantities to be estimated: $\sigma_W(v^{\text{R}})$, $\sigma_W(v^{\text{T}})$ and $\sigma_W(v^{\text{N}})$, i.e., the diffusion coefficients in the radial, transversal, and normal components of the velocity. In this case, the NHF is run with $N_1=100$ and $N_2=100$ samples. Table \ref{tab:Algo_perf3} shows the RMSEs and run-times of the algorithms using this set-up.

\begin{figure}[H]
    \hspace{-1.7cm}
    \includegraphics[scale=0.36]{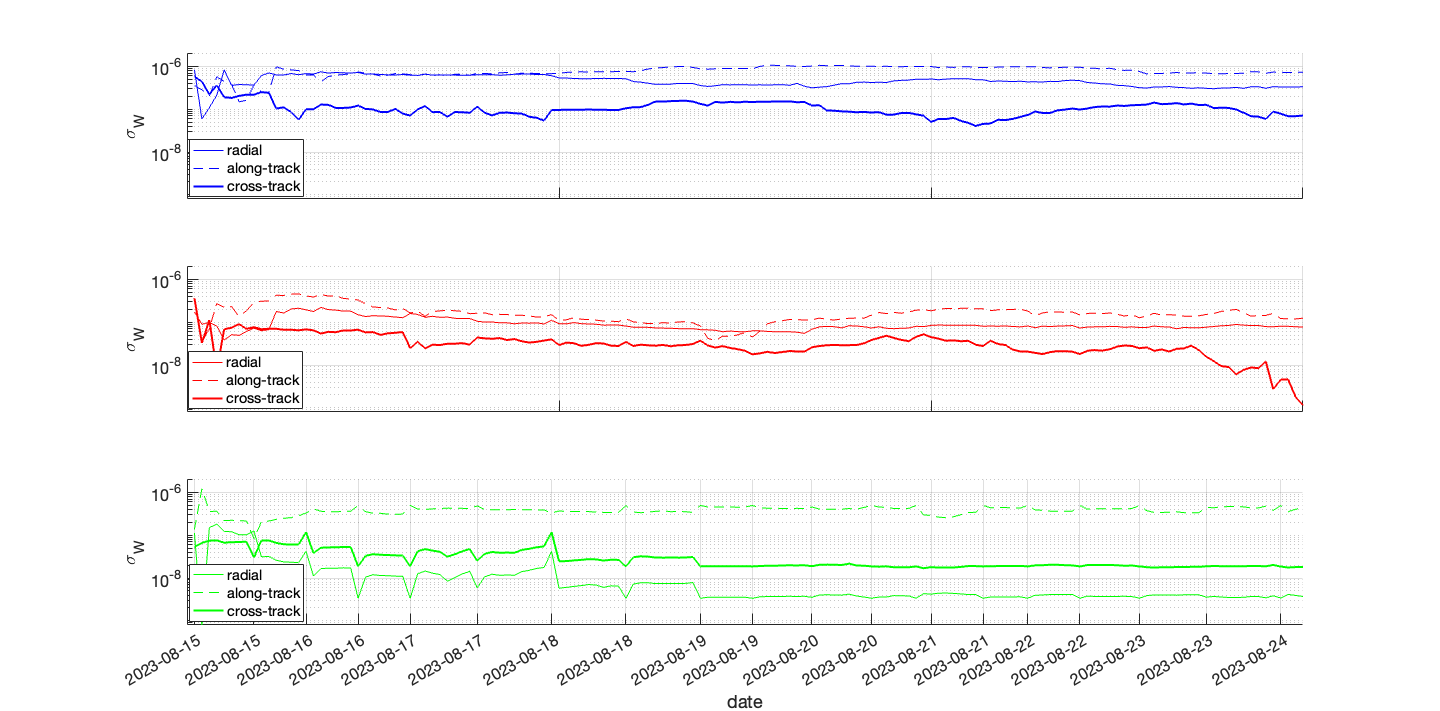}
    \caption{ In this figure, the $\bar\sigma_W$ estimated by the EnKFup (in blue), the EnPFup (in red) and the NHF (in green) algorithms , in the presence of a simplified dynamical model. The thick solid line shows the cross-track component, the thin solid line shows the radial component, and the dashed line shows the along-track component.}
    \label{fig:3sigma_rtn}
\end{figure}

From Table \ref{tab:Algo_perf3}, it can be seen that the estimation performance does not improve compared to Table \ref{tab:Algo_perf2}. Despite working with more degrees of freedom in the parameter space, the fact that more parameter components need to be adjusted causes uncertainty to increase. The NHF is the algorithm with the largest errors, whilst the EnPFup seems to be the most stable of the three.
Figure \ref{fig:3sigma_rtn} shows the estimated RTN components in $\bar\sigma_W$ for the EnKFup, EnPFup and NHF.

\begin{table}[h]
\caption{Performance comparison for the three algorithms. The metrics are the RMSE in position and the RMSE in velocity of the estimated trajectory compared to the reference trajectory, the standard deviations of the obtained errors, and the run-time.} 
\normalsize
\begin{center}
    \hspace*{0.1cm}\begin{tabular}{l c c c}
    Algorithm & Position RMSE (m) & Velocity RMSE (m/s) & Run-time (min)\\
    \hline 
    EnKFup & $86.90 \pm 11.24$ & $0.12 \pm 0.2$ & $16.31$\\ 
    EnPFup & $73.84 \pm 11.22$ & $0.09 \pm 0.03$ & $15.19$\\
    NHF & $ 111.75 \pm 25.18$ & $0.15 \pm 0.06$ & $243.51$ \\
    \hline 
    \end{tabular}
    \label{tab:Algo_perf3}
\end{center}
\end{table}

From Figure \ref{fig:3sigma_rtn}, it is shown that both the EnKFup and EnPFup estimate the along-track component to be the largest and the cross-track component to be the smallest. The NHF shows agreement on the along-track component, but estimates the radial component to be the smallest.
Note that the truncation of certain acceleration terms in the LF dynamical model may induce error in particular directions. LEO trajectories typically show a larger amount of uncertainty in the along-track component during tracking, while the cross-track and radial components typically show lower levels of error. The interchangeability between the latter two, as shown by the algorithms in Figure \ref{fig:3sigma_rtn} may arise due to variations in the magnitudes of the truncated accelerations within the simplified dynamical model.

\section{Conclusions}\label{sec:CONC}
We present three recursive filters, the EnKFup, the EnPFup and the NHF, which incorporate notions of PFs and Gaussian filters, to track a LEO spacecraft for a period of $\sim 9$ days. By modeling orbital dynamics as an SDE, we are able to simultaneously track the state (position and velocity) of a spacecraft and estimate the diffusion coefficient magnitude in the SDE, i.e., the process noise magnitude. 

In order to validate the proposed algorithms, two scenarios are tested. The first uses a set of measurements computed from an HF stochastically propagated reference orbit with a given diffusion coefficient $\sigma_W$. To assess whether the algorithms are capable of correctly estimating the parameter, these are run with the same dynamical model with the aim of estimating $\sigma_W$.
The second scenario uses measurements computed from an HF deterministic reference orbit. In order to assess whether the algorithms are capable of performing well while using a very simplified dynamical model, these are run on a stochastic LF model, and the performance of stochastic parametrization is assessed, i.e., determining how well the unknown accelerations in the dynamical model are accounted for by estimating the parameter $\sigma_W$.

In the first test, the three algorithms achieve low errors in both position and velocity compared to an NBLS method, and can estimate the magnitude of the nominal parameter $\sigma_W$ accurately, with the NHF achieving the lowest estimation errors in the parameter space, but showing the highest computational cost. The EnPFup is chosen over the EnKFup due to its lower errors in position, but similar runtime. For the second test, all three algorithms achieve low errors in position, and velocity so in terms of computational cost, the EnPFup is determined to achieve the best trade-off between cost and accuracy out of the three algorithms. 
The algorithms are therefore able to track a spacecraft in the realistic case where there are significant sources of uncertainty, and/or the model used is simplified in order to cut run-time costs. The result is an ability to ``estimate what you do not know" online, and appropriately characterize the uncertainty of the system, whilst performing satisfactory tracking of a spacecraft in the presence of observations.

\section*{Acknowledgements}

This work has been supported by Comunidad de Madrid (project ref. IND2020/TIC-17539), {\em Agencia Estatal de Investigaci\'on} (ref. PID2021-125159NB-I00 TYCHE) funded by MCIN/AEI/10.13039/501100011033 and by ``ERDF A way of making Europe", and the Office of Naval Research (award N00014-22-1-2647).

\section*{Declarations}
The authors declare that they have no competing financial interests that could influence the work reported in this paper.





\bibliography{Sequential_Filtering_Techniques_for_Simultaneous_Tracking_and_Parameter_Estimation}


\begin{thebibliography}{48}
\ifx \bisbn   \undefined \def \bisbn  #1{ISBN #1}\fi
\ifx \binits  \undefined \def \binits#1{#1}\fi
\ifx \bauthor  \undefined \def \bauthor#1{#1}\fi
\ifx \batitle  \undefined \def \batitle#1{#1}\fi
\ifx \bjtitle  \undefined \def \bjtitle#1{#1}\fi
\ifx \bvolume  \undefined \def \bvolume#1{\textbf{#1}}\fi
\ifx \byear  \undefined \def \byear#1{#1}\fi
\ifx \bissue  \undefined \def \bissue#1{#1}\fi
\ifx \bfpage  \undefined \def \bfpage#1{#1}\fi
\ifx \blpage  \undefined \def \blpage #1{#1}\fi
\ifx \burl  \undefined \def \burl#1{\textsf{#1}}\fi
\ifx \doiurl  \undefined \def \doiurl#1{\url{https://doi.org/#1}}\fi
\ifx \betal  \undefined \def \betal{\textit{et al.}}\fi
\ifx \binstitute  \undefined \def \binstitute#1{#1}\fi
\ifx \binstitutionaled  \undefined \def \binstitutionaled#1{#1}\fi
\ifx \bctitle  \undefined \def \bctitle#1{#1}\fi
\ifx \beditor  \undefined \def \beditor#1{#1}\fi
\ifx \bpublisher  \undefined \def \bpublisher#1{#1}\fi
\ifx \bbtitle  \undefined \def \bbtitle#1{#1}\fi
\ifx \bedition  \undefined \def \bedition#1{#1}\fi
\ifx \bseriesno  \undefined \def \bseriesno#1{#1}\fi
\ifx \blocation  \undefined \def \blocation#1{#1}\fi
\ifx \bsertitle  \undefined \def \bsertitle#1{#1}\fi
\ifx \bsnm \undefined \def \bsnm#1{#1}\fi
\ifx \bsuffix \undefined \def \bsuffix#1{#1}\fi
\ifx \bparticle \undefined \def \bparticle#1{#1}\fi
\ifx \barticle \undefined \def \barticle#1{#1}\fi
\bibcommenthead
\ifx \bconfdate \undefined \def \bconfdate #1{#1}\fi
\ifx \botherref \undefined \def \botherref #1{#1}\fi
\ifx \url \undefined \def \url#1{\textsf{#1}}\fi
\ifx \bchapter \undefined \def \bchapter#1{#1}\fi
\ifx \bbook \undefined \def \bbook#1{#1}\fi
\ifx \bcomment \undefined \def \bcomment#1{#1}\fi
\ifx \oauthor \undefined \def \oauthor#1{#1}\fi
\ifx \citeauthoryear \undefined \def \citeauthoryear#1{#1}\fi
\ifx \endbibitem  \undefined \def \endbibitem {}\fi
\ifx \bconflocation  \undefined \def \bconflocation#1{#1}\fi
\ifx \arxivurl  \undefined \def \arxivurl#1{\textsf{#1}}\fi
\csname PreBibitemsHook\endcsname

\bibitem[\protect\citeauthoryear{{ESA Space Debris Office}}{2023}]{ESAdata}
\begin{botherref}
\oauthor{\bsnm{{ESA Space Debris Office}}}:
Space Debris Environment Report, 7th edition.
Space Environment Statistics
(2023)
\end{botherref}
\endbibitem

\bibitem[\protect\citeauthoryear{Gelb}{1974}]{Gelb}
\begin{bbook}
\bauthor{\bsnm{Gelb}, \binits{A.}}:
\bbtitle{Applied Optimal Estimation}.
\bpublisher{MIT Press},
\blocation{Cambridge, MA}
(\byear{1974})
\end{bbook}
\endbibitem

\bibitem[\protect\citeauthoryear{Maybeck}{1979}]{Maybeck}
\begin{bbook}
\bauthor{\bsnm{Maybeck}, \binits{P.S.}}:
\bbtitle{Stochastic Models, Estimation, and Control, Volume 1}.
\bpublisher{Academic Press},
\blocation{New York, NY}
(\byear{1979})
\end{bbook}
\endbibitem

\bibitem[\protect\citeauthoryear{Fraser}{1969}]{Fraser}
\begin{bbook}
\bauthor{\bsnm{Fraser}, \binits{G.E.}}:
\bbtitle{The Estimation of Dynamic Systems}.
\bpublisher{Prentice-Hall},
\blocation{Englewood Cliffs, NJ}
(\byear{1969})
\end{bbook}
\endbibitem

\bibitem[\protect\citeauthoryear{Patil and T}{2013}]{Patil}
\begin{barticle}
\bauthor{\bsnm{Patil}, \binits{P.}},
\bauthor{\bsnm{T}, \binits{S.K.}}:
\batitle{Orbit determination using batch sequential filter}.
\bjtitle{International Journal of Engineering Research and Technology}
\bvolume{1}(\bissue{4}),
\bfpage{1}--\blpage{5}
(\byear{2013})
\end{barticle}
\endbibitem

\bibitem[\protect\citeauthoryear{Kalman}{1960}]{Kalman}
\begin{barticle}
\bauthor{\bsnm{Kalman}, \binits{R.E.}}:
\batitle{A new approach to linear filtering and prediction problems}.
\bjtitle{Transactions of the ASME—Journal of Basic Engineering}
\bvolume{82},
\bfpage{35}--\blpage{45}
(\byear{1960})
\doiurl{10.1115/1.3662552}
\end{barticle}
\endbibitem

\bibitem[\protect\citeauthoryear{Gordon et~al.}{1993}]{Gordon}
\begin{barticle}
\bauthor{\bsnm{Gordon}, \binits{N.J.}},
\bauthor{\bsnm{Salmond}, \binits{D.J.}},
\bauthor{\bsnm{Smith}, \binits{A.F.M.}}:
\batitle{Novel approach to nonlinear/non-{G}aussian {B}ayesian state estimation}.
\bjtitle{IEE Proceedings F - Radar and Signal Processing}
\bvolume{140}(\bissue{2}),
\bfpage{107}--\blpage{113}
(\byear{1993})
\doiurl{10.1049/ip-f-2.1993.0015}
\end{barticle}
\endbibitem

\bibitem[\protect\citeauthoryear{Alspach and Sorenson}{1972}]{Alspach}
\begin{barticle}
\bauthor{\bsnm{Alspach}, \binits{D.L.}},
\bauthor{\bsnm{Sorenson}, \binits{H.W.}}:
\batitle{Nonlinear {B}ayesian estimation using {G}aussian sum approximations}.
\bjtitle{IEEE Transactions on Automatic Control}
\bvolume{17}(\bissue{4}),
\bfpage{439}--\blpage{448}
(\byear{1972})
\doiurl{10.1109/TAC.1972.1100103}
\end{barticle}
\endbibitem

\bibitem[\protect\citeauthoryear{Doucet et~al.}{2000}]{Doucet2000}
\begin{barticle}
\bauthor{\bsnm{Doucet}, \binits{A.}},
\bauthor{\bsnm{Godsill}, \binits{S.}},
\bauthor{\bsnm{Andrieu}, \binits{C.}}:
\batitle{On sequential {M}onte {C}arlo sampling methods for {B}ayesian filtering}.
\bjtitle{Statistics and Computing}
\bvolume{10},
\bfpage{197}--\blpage{208}
(\byear{2000})
\doiurl{10.1023/A:1008935410038}
\end{barticle}
\endbibitem

\bibitem[\protect\citeauthoryear{Anderson and Moore}{1979}]{Anderson}
\begin{bbook}
\bauthor{\bsnm{Anderson}, \binits{B.D.O.}},
\bauthor{\bsnm{Moore}, \binits{J.B.}}:
\bbtitle{Optimal Filtering}.
\bsertitle{Prentice-Hall Information and System Sciences Series}.
\bpublisher{Prentice-Hall},
\blocation{Englewood Cliffs, NJ}
(\byear{1979})
\end{bbook}
\endbibitem

\bibitem[\protect\citeauthoryear{Julier and Uhlmann}{1995}]{Julier}
\begin{barticle}
\bauthor{\bsnm{Julier}, \binits{S.J.}},
\bauthor{\bsnm{Uhlmann}, \binits{J.K.}}:
\batitle{The unscented {K}alman filter for nonlinear estimation}.
\bjtitle{Proceedings of the 1995 IEEE Aerospace Conference}
\bvolume{6},
\bfpage{3}--\blpage{9}
(\byear{1995})
\doiurl{10.1109/AERO.1995.498946}
\end{barticle}
\endbibitem

\bibitem[\protect\citeauthoryear{Arasaratnam and Haykin}{2009}]{ckf}
\begin{barticle}
\bauthor{\bsnm{Arasaratnam}, \binits{H.}},
\bauthor{\bsnm{Haykin}, \binits{S.}}:
\batitle{Cubature kalman filters}.
\bjtitle{IEEE Transactions on Automatic Control}
\bvolume{54}(\bissue{6}),
\bfpage{1254}--\blpage{1269}
(\byear{2009})
\end{barticle}
\endbibitem

\bibitem[\protect\citeauthoryear{Segan}{2017}]{Segan}
\begin{barticle}
\bauthor{\bsnm{Segan}, \binits{S.}}:
\batitle{Orbit determination and parameter estimation: Extended {K}alman filter ({EKF}) versus least squares orbit determination}.
\bjtitle{Celestial Mechanics and Dynamical Astronomy}
\bvolume{129},
\bfpage{345}--\blpage{368}
(\byear{2017})
\end{barticle}
\endbibitem

\bibitem[\protect\citeauthoryear{Li}{2021}]{5ckf}
\begin{barticle}
\bauthor{\bsnm{Li}, \binits{Z.}}:
\batitle{A novel fifth-degree cubature {K}alman filter for real-time orbit determination by radar}.
\bjtitle{Aerospace Science and Technology}
\bvolume{89},
\bfpage{12}--\blpage{20}
(\byear{2021})
\end{barticle}
\endbibitem

\bibitem[\protect\citeauthoryear{Doucet et~al.}{2001}]{Doucet2001}
\begin{bbook}
\bauthor{\bsnm{Doucet}, \binits{A.}},
\bauthor{\bsnm{Freitas}, \binits{N.}},
\bauthor{\bsnm{Gordon}, \binits{N.}}:
\bbtitle{Sequential {M}onte {C}arlo Methods in Practice}.
\bsertitle{Statistics for Engineering and Information Science}.
\bpublisher{Springer},
\blocation{New York, NY}
(\byear{2001}).
\doiurl{10.1007/978-1-4757-3437-9}
\end{bbook}
\endbibitem

\bibitem[\protect\citeauthoryear{Djuric et~al.}{2003}]{Djuric2003}
\begin{barticle}
\bauthor{\bsnm{Djuric}, \binits{P.M.}},
\bauthor{\bsnm{Kotecha}, \binits{J.H.}},
\bauthor{\bsnm{Zhang}, \binits{J.}},
\bauthor{\bsnm{Huang}, \binits{Y.}},
\bauthor{\bsnm{Bugallo}, \binits{M.F.}},
\bauthor{\bsnm{Miguez}, \binits{J.}}:
\batitle{Particle filtering}.
\bjtitle{IEEE Signal Processing Magazine}
\bvolume{20}(\bissue{5}),
\bfpage{19}--\blpage{38}
(\byear{2003})
\doiurl{10.1109/MSP.2003.1233220}
\end{barticle}
\endbibitem

\bibitem[\protect\citeauthoryear{Cappé et~al.}{2007}]{Cappe}
\begin{botherref}
\oauthor{\bsnm{Cappé}, \binits{O.}},
\oauthor{\bsnm{Godsill}, \binits{S.J.}},
\oauthor{\bsnm{Moulines}, \binits{E.}}:
An overview of existing methods and recent advances in sequential {M}onte {C}arlo.
Proceedings of the IEEE
\textbf{95}(5)
(2007)
\end{botherref}
\endbibitem

\bibitem[\protect\citeauthoryear{Kitagawa}{1996}]{Kitagawa}
\begin{barticle}
\bauthor{\bsnm{Kitagawa}, \binits{G.}}:
\batitle{{M}onte {C}arlo filter and smoother for non-{G}aussian nonlinear time series}.
\bjtitle{Journal of Computational and Graphical Statistics}
\bvolume{5}(\bissue{1}),
\bfpage{1}--\blpage{25}
(\byear{1996})
\doiurl{10.1080/10618600.1996.10474713}
\end{barticle}
\endbibitem

\bibitem[\protect\citeauthoryear{Maskell et~al.}{2002}]{Maskell}
\begin{bchapter}
\bauthor{\bsnm{Maskell}, \binits{S.}},
\bauthor{\bsnm{Briers}, \binits{M.}},
\bauthor{\bsnm{Wright}, \binits{R.}},
\bauthor{\bsnm{Horridge}, \binits{P.}}:
\bctitle{Tracking using a radar and a problem specific proposal distribution in a particle filter}.
In: \bbtitle{Proceedings of the Fifth International Conference on Information Fusion},
vol. \bseriesno{2},
pp. \bfpage{867}--\blpage{872}
(\byear{2002}).
\bcomment{IEEE}
\end{bchapter}
\endbibitem

\bibitem[\protect\citeauthoryear{Li et~al.}{2015}]{Djuric15}
\begin{botherref}
\oauthor{\bsnm{Li}, \binits{T.}},
\oauthor{\bsnm{Bolic}, \binits{M.}},
\oauthor{\bsnm{Djuric}, \binits{P.M.}}:
Resampling methods for particle filtering.
IEEE Signal Processing Magazine,
70--86
(2015)
\end{botherref}
\endbibitem

\bibitem[\protect\citeauthoryear{Snyder et~al.}{2008}]{Snyder08}
\begin{barticle}
\bauthor{\bsnm{Snyder}, \binits{C.}},
\bauthor{\bsnm{Bengtsson}, \binits{T.}},
\bauthor{\bsnm{Bickel}, \binits{P.}},
\bauthor{\bsnm{Anderson}, \binits{J.L.}}:
\batitle{Obstacles to high-dimensional particle filtering}.
\bjtitle{Monthly Weather Review}
\bvolume{136}(\bissue{12}),
\bfpage{4629}--\blpage{4640}
(\byear{2008})
\doiurl{10.1175/2008MWR2529.1}
\end{barticle}
\endbibitem

\bibitem[\protect\citeauthoryear{Pardal}{2021}]{Pardal1}
\begin{barticle}
\bauthor{\bsnm{Pardal}, \binits{P.C.P.M.}}:
\batitle{The particle filter sample impoverishment problem in the orbit determination application}.
\bjtitle{Aerospace Science and Technology}
\bvolume{104},
\bfpage{20}--\blpage{32}
(\byear{2021})
\end{barticle}
\endbibitem

\bibitem[\protect\citeauthoryear{McCabe}{2016}]{McCabe}
\begin{barticle}
\bauthor{\bsnm{McCabe}, \binits{J.S.}}:
\batitle{Particle filter methods for space object tracking}.
\bjtitle{Acta Astronautica}
\bvolume{125},
\bfpage{50}--\blpage{62}
(\byear{2016})
\end{barticle}
\endbibitem

\bibitem[\protect\citeauthoryear{Mashiku}{2022}]{Mashiku}
\begin{barticle}
\bauthor{\bsnm{Mashiku}, \binits{A.}}:
\batitle{Statistical orbit determination using the particle filter for incorporating non-{G}aussian uncertainties}.
\bjtitle{Acta Astronautica}
\bvolume{150},
\bfpage{129}--\blpage{140}
(\byear{2022})
\end{barticle}
\endbibitem

\bibitem[\protect\citeauthoryear{Escribano}{2022}]{Escribano}
\begin{barticle}
\bauthor{\bsnm{Escribano}, \binits{G.}}:
\batitle{Automatic maneuver detection and tracking of space objects in optical survey scenarios based on stochastic hybrid systems formulation}.
\bjtitle{Advances in Space Research}
\bvolume{68}(\bissue{10}),
\bfpage{3156}--\blpage{3168}
(\byear{2022})
\end{barticle}
\endbibitem

\bibitem[\protect\citeauthoryear{Evensen}{1996}]{Evensen}
\begin{barticle}
\bauthor{\bsnm{Evensen}, \binits{G.}}:
\batitle{The ensemble kalman filter: Theoretical formulation and practical implementation}.
\bjtitle{Ocean Dynamics}
\bvolume{53}(\bissue{4}),
\bfpage{343}--\blpage{367}
(\byear{1996})
\doiurl{10.1007/s10236-003-0036-9}
\end{barticle}
\endbibitem

\bibitem[\protect\citeauthoryear{Gamper et~al.}{2019}]{Gamper}
\begin{bchapter}
\bauthor{\bsnm{Gamper}, \binits{E.}},
\bauthor{\bsnm{Kebschull}, \binits{C.}},
\bauthor{\bsnm{Stoll}, \binits{E.}}:
\bctitle{Statistical orbit determination using the ensemble {K}alman filter}.
In: \bbtitle{1st NEO and Debris Detection Conference}
(\byear{2019}).
\burl{https://conference.sdo.esoc.esa.int/proceedings/neosst1/paper/458}
\end{bchapter}
\endbibitem

\bibitem[\protect\citeauthoryear{DeMars}{2014}]{DeMars_GMM}
\begin{barticle}
\bauthor{\bsnm{DeMars}, \binits{K.}}:
\batitle{Probabilistic initial orbit determination using {G}aussian mixture models}.
\bjtitle{Celestial Mechanics and Dynamical Astronomy}
\bvolume{118}(\bissue{3}),
\bfpage{171}--\blpage{184}
(\byear{2014})
\end{barticle}
\endbibitem

\bibitem[\protect\citeauthoryear{Yun}{2019}]{Yun1}
\begin{barticle}
\bauthor{\bsnm{Yun}, \binits{S.}}:
\batitle{Kernel-based ensemble {G}aussian mixture filtering for orbit determination with sparse data}.
\bjtitle{IEEE Transactions on Aerospace and Electronic Systems}
\bvolume{55}(\bissue{2}),
\bfpage{845}--\blpage{858}
(\byear{2019})
\end{barticle}
\endbibitem

\bibitem[\protect\citeauthoryear{Raihan and Chakravorty}{2021}]{Chakravorty2}
\begin{barticle}
\bauthor{\bsnm{Raihan}, \binits{D.}},
\bauthor{\bsnm{Chakravorty}, \binits{S.}}:
\batitle{A {UKF-PF} based hybrid estimation scheme for space object tracking}.
\bjtitle{Journal of Guidance, Control, and Dynamics}
\bvolume{44}(\bissue{8}),
\bfpage{1456}--\blpage{1470}
(\byear{2021})
\end{barticle}
\endbibitem

\bibitem[\protect\citeauthoryear{Zhang et~al.}{2020}]{zhang2020identification}
\begin{barticle}
\bauthor{\bsnm{Zhang}, \binits{L.}},
\bauthor{\bsnm{Sidoti}, \binits{D.}},
\bauthor{\bsnm{Bienkowski}, \binits{A.}},
\bauthor{\bsnm{Pattipati}, \binits{K.R.}},
\bauthor{\bsnm{Bar-Shalom}, \binits{Y.}},
\bauthor{\bsnm{Kleinman}, \binits{D.L.}}:
\batitle{On the identification of noise covariances and adaptive {K}alman filtering: A new look at a 50 year-old problem}.
\bjtitle{IEEE Access}
\bvolume{8},
\bfpage{59362}--\blpage{59388}
(\byear{2020})
\end{barticle}
\endbibitem

\bibitem[\protect\citeauthoryear{Dun{\'\i}k et~al.}{2017}]{dunik2017noise}
\begin{barticle}
\bauthor{\bsnm{Dun{\'\i}k}, \binits{J.}},
\bauthor{\bsnm{Straka}, \binits{O.}},
\bauthor{\bsnm{Kost}, \binits{O.}},
\bauthor{\bsnm{Havl{\'\i}k}, \binits{J.}}:
\batitle{Noise covariance matrices in state-space models: A survey and comparison of estimation methods—part {I}}.
\bjtitle{International Journal of Adaptive Control and Signal Processing}
\bvolume{31}(\bissue{11}),
\bfpage{1505}--\blpage{1543}
(\byear{2017})
\end{barticle}
\endbibitem

\bibitem[\protect\citeauthoryear{on~Covariance~Realism}{2016}]{poore2016covariance}
\begin{botherref}
\oauthor{\bsnm{Covariance~Realism}, \binits{W.G.}}:
Covariance and uncertainty realism in space surveillance and tracking.
Technical report,
Astrodynamics Innovation Committee
(2016)
\end{botherref}
\endbibitem

\bibitem[\protect\citeauthoryear{Stacey et~al.}{2020}]{Stacey1}
\begin{botherref}
\oauthor{\bsnm{Stacey}, \binits{N.}},
\oauthor{\bsnm{Furfaro}, \binits{R.}},
\oauthor{\bsnm{Fossum}, \binits{E.J.}}:
Time-Adaptive Process Noise Estimation for Orbit Determination.
Available at: \url{https://arxiv.org/abs/2001.03273}
(2020)
\end{botherref}
\endbibitem

\bibitem[\protect\citeauthoryear{Stacey et~al.}{2019}]{Stacey2}
\begin{botherref}
\oauthor{\bsnm{Stacey}, \binits{N.}},
\oauthor{\bsnm{Furfaro}, \binits{R.}},
\oauthor{\bsnm{Fossum}, \binits{E.J.}}:
Adaptive and Dynamically Constrained Process Noise Estimation for Orbit Determination.
Available at: \url{https://arxiv.org/abs/1909.07921}
(2019)
\end{botherref}
\endbibitem

\bibitem[\protect\citeauthoryear{Cano et~al.}{2023}]{Cano1}
\begin{barticle}
\bauthor{\bsnm{Cano}, \binits{A.}},
\bauthor{\bsnm{Pastor}, \binits{A.}},
\bauthor{\bsnm{Escobar}, \binits{D.}},
\bauthor{\bsnm{Míguez}, \binits{J.}},
\bauthor{\bsnm{Sanjurjo-Rivo}, \binits{M.}}:
\batitle{Covariance determination for improving uncertainty realism in orbit determination and propagation}.
\bjtitle{Advances in Space Research}
\bvolume{72},
\bfpage{2759}--\blpage{2777}
(\byear{2023})
\doiurl{10.1016/j.asr.2023.06.010}
\end{barticle}
\endbibitem

\bibitem[\protect\citeauthoryear{Cano et~al.}{2022}]{Cano2}
\begin{barticle}
\bauthor{\bsnm{Cano}, \binits{A.}},
\bauthor{\bsnm{Pastor}, \binits{A.}},
\bauthor{\bsnm{Fernandez}, \binits{S.}},
\bauthor{\bsnm{M\'{i}guez}, \binits{J.}},
\bauthor{\bsnm{Sanjurjo-Rivo}, \binits{M.}},
\bauthor{\bsnm{Escobar}, \binits{D.}}:
\batitle{Improving orbital uncertainty realism through covariance determination in {GEO}}.
\bjtitle{The Journal of the Astronautical Sciences}
\bvolume{69}(\bissue{5}),
\bfpage{1394}--\blpage{1420}
(\byear{2022})
\doiurl{10.1007/s40295-022-00343-x}
\end{barticle}
\endbibitem

\bibitem[\protect\citeauthoryear{Luo and Yang}{2017}]{luo}
\begin{barticle}
\bauthor{\bsnm{Luo}, \binits{Y.-z.}},
\bauthor{\bsnm{Yang}, \binits{Z.}}:
\batitle{A review of uncertainty propagation in orbital mechanics}.
\bjtitle{Progress in Aerospace Sciences}
\bvolume{89},
\bfpage{23}--\blpage{39}
(\byear{2017})
\end{barticle}
\endbibitem

\bibitem[\protect\citeauthoryear{Pérez-Vieites and Míguez}{2021}]{PerezVieites2021}
\begin{botherref}
\oauthor{\bsnm{Pérez-Vieites}, \binits{S.}},
\oauthor{\bsnm{Míguez}, \binits{J.}}:
Nested {G}aussian filters for recursive {B}ayesian inference and nonlinear tracking in state space models.
Signal Processing
\textbf{189}
(2021)
\doiurl{10.1016/j.sigpro.2021.108295}
\end{botherref}
\endbibitem

\bibitem[\protect\citeauthoryear{Rümelin}{1982}]{Rumelin1982}
\begin{barticle}
\bauthor{\bsnm{Rümelin}, \binits{W.}}:
\batitle{Numerical treatment of stochastic differential equations}.
\bjtitle{SIAM Journal on Numerical Analysis}
\bvolume{19}(\bissue{3}),
\bfpage{604}--\blpage{613}
(\byear{1982})
\doiurl{10.1137/0719040}
\end{barticle}
\endbibitem

\bibitem[\protect\citeauthoryear{Kloeden and Platen}{1992}]{Kloeden}
\begin{bbook}
\bauthor{\bsnm{Kloeden}, \binits{P.E.}},
\bauthor{\bsnm{Platen}, \binits{E.}}:
\bbtitle{Numerical Solution of Stochastic Differential Equations}.
\bsertitle{Stochastic Modelling and Applied Probability},
vol. \bseriesno{23}.
\bpublisher{Springer},
\blocation{Berlin, Germany}
(\byear{1992}).
\doiurl{10.1007/978-3-662-12616-5}
\end{bbook}
\endbibitem

\bibitem[\protect\citeauthoryear{Bishop and Moral}{2023}]{Bishopp}
\begin{barticle}
\bauthor{\bsnm{Bishop}, \binits{A.N.}},
\bauthor{\bsnm{Moral}, \binits{P.D.}}:
\batitle{On the mathematical theory of ensemble (linear-{G}aussian) {K}alman–{B}ucy filtering}.
\bjtitle{Mathematics of Control, Signals, and Systems}
\bvolume{35},
\bfpage{1}--\blpage{83}
(\byear{2023})
\doiurl{10.1007/s00498-023-00357-2}
\end{barticle}
\endbibitem

\bibitem[\protect\citeauthoryear{Pérez-Vieites et~al.}{2018}]{Perez-vieites_enkf}
\begin{botherref}
\oauthor{\bsnm{Pérez-Vieites}, \binits{S.}},
\oauthor{\bsnm{Mariño}, \binits{I.P.}},
\oauthor{\bsnm{Míguez}, \binits{J.}}:
A probabilistic scheme for joint parameter estimation and state prediction in complex dynamical systems.
Physical Review E
\textbf{98}(6)
(2018)
\doiurl{10.1103/PhysRevE.98.063305}
\end{botherref}
\endbibitem

\bibitem[\protect\citeauthoryear{Elvira and Miguez}{2021}]{Elvira}
\begin{botherref}
\oauthor{\bsnm{Elvira}, \binits{V.}},
\oauthor{\bsnm{Miguez}, \binits{J.}}:
On the performance of particle filters with adaptive number of particles.
Statistics and Computing
\textbf{31}(81)
(2021)
\end{botherref}
\endbibitem

\bibitem[\protect\citeauthoryear{Crisan and M{\'\i}guez}{2017a}]{crisan1}
\begin{barticle}
\bauthor{\bsnm{Crisan}, \binits{D.}},
\bauthor{\bsnm{M{\'\i}guez}, \binits{J.}}:
\batitle{Nested particle filters for online parameter estimation in discrete-time state-space {M}arkov models}.
\bjtitle{Bernoulli}
\bvolume{23}(\bissue{4A}),
\bfpage{2672}--\blpage{2714}
(\byear{2017})
\end{barticle}
\endbibitem

\bibitem[\protect\citeauthoryear{Crisan and M{\'\i}guez}{2017b}]{crisan2}
\begin{barticle}
\bauthor{\bsnm{Crisan}, \binits{D.}},
\bauthor{\bsnm{M{\'\i}guez}, \binits{J.}}:
\batitle{Uniform convergence over time of a nested particle filtering scheme for recursive parameter estimation in state-space {M}arkov models}.
\bjtitle{Advances in Applied Probability}
\bvolume{49}(\bissue{4}),
\bfpage{1148}--\blpage{1175}
(\byear{2017})
\end{barticle}
\endbibitem

\bibitem[\protect\citeauthoryear{Mahooti}{2024}]{Mahooti}
\begin{botherref}
\oauthor{\bsnm{Mahooti}, \binits{M.}}:
High precision orbit propagator.
MATLAB Central File Exchange
(2024)
\end{botherref}
\endbibitem

\bibitem[\protect\citeauthoryear{Petropoulos}{2005}]{Qlaw}
\begin{barticle}
\bauthor{\bsnm{Petropoulos}, \binits{A.}}:
\batitle{Refinements to the {Q}-law for the low-thrust orbit transfers}.
\bjtitle{Jet Propulsion Laboratory, NASA}
(\byear{2005})
\doiurl{10.1016/j.asr.2023.06.010}
\end{barticle}
\endbibitem

\end{thebibliography}

\end{document}